\input harvmac
\input epsf

\def\ndt{\noindent}

\def\Ga{{\Gamma}}

\def\K3{{\bf K3}}
\def\journal#1&#2(#3){\unskip, \sl #1\ \bf #2 \rm(19#3) }
\def\andjournal#1&#2(#3){\sl #1~\bf #2 \rm (19#3) }

\def\bar{\overline}

\def\tilde{\widetilde}

\def\frac#1#2{{#1\over#2}}

\def\vev#1{\langle#1\rangle}

\def\inbar{\,\vrule height1.5ex width.4pt depth0pt}
\def\IC{\relax\hbox{$\inbar\kern-.3em{\rm C}$}}
\def\IR{\relax{\rm I\kern-.18em R}}
\def\IP{\relax{\rm I\kern-.18em P}}

%
%


%
\catcode`\@=11
\def\slash#1{\mathord{\mathpalette\c@ncel{#1}}}
\overfullrule=0pt

\def\CC{{\cal C}}

\def\GG{{\cal G}}

\def\OO{{\cal O}}

\def\ZZ{{\cal Z}}

\def\underrel#1\over#2{\mathrel{\mathop{\kern\z@#1}\limits_{#2}}}

\catcode`\@=12


%

\def\vev#1{\left\langle #1 \right\rangle}

\def\exp{{\rm exp}}


\def\IL{\relax{\rm I\kern-.18em L}}
\def\IH{\relax{\rm I\kern-.18em H}}
\def\IR{\relax{\rm I\kern-.18em R}}
\def\IC{\relax\hbox{$\inbar\kern-.3em{\rm C}$}}
\def\IZ{{\bf Z}}





\def\makeblankbox#1#2{\hbox{\lower\dp0\vbox{\hidehrule{#1}{#2}%
   \kern -#1
   \hbox to \wd0{\hidevrule{#1}{#2}%
      \raise\ht0\vbox to #1{}
      \lower\dp0\vtop to #1{}
      \hfil\hidevrule{#2}{#1}}%
   \kern-#1\hidehrule{#2}{#1}}}%
}%
\def\hidehrule#1#2{\kern-#1\hrule height#1 depth#2 \kern-#2}%
\def\hidevrule#1#2{\kern-#1{\dimen0=#1\advance\dimen0 by #2\vrule
    width\dimen0}\kern-#2}%
\def\openbox{\ht0=1.2mm \dp0=1.2mm \wd0=2.4mm  \raise 2.75pt
\makeblankbox {.25pt} {.25pt}  }

\def\bun#1/#2{\leavevmode
   \kern.1em \raise .5ex \hbox{\the\scriptfont0 #1}%
   \kern-.1em $/$%
   \kern-.15em \lower .25ex \hbox{\the\scriptfont0 #2}%
}

\def\opensquare{\ht0=3.4mm \dp0=3.4mm \wd0=6.8mm  \raise 2.7pt
\makeblankbox {.25pt} {.25pt}  }


\def\sector#1#2{\ {\scriptstyle #1}\hskip 1mm
\mathop{\opensquare}\limits_{\lower
1mm\hbox{$\scriptstyle#2$}}\hskip 1mm}

\def\tsector#1#2{\ {\scriptstyle #1}\hskip 1mm
\mathop{\opensquare}\limits_{\lower
1mm\hbox{$\scriptstyle#2$}}^\sim\hskip 1mm}

\def \apr {\alpha'}

\def \ov {\over}
\def \p {\partial}
\def \ha {{1 \ov 2}}
\def\le{\left}
\def\ri{\right}

\def\IZ{{\bf Z}}

\def \om {\omega}
\def \ep {\epsilon}

\def\th{\theta}

\def\Ga{\Gamma}

\def\cB{{\cal B}}
\def\bcB{{\bar{\cal B}}}

\def\rw{u} 

\lref\JimboSS{ M.~Jimbo and T.~Miwa, ``QKZ equation with $|$q$|$=1
and correlation functions of the XXZ model in the gapless
regime,'' J.\ Phys.\ A {\bf 29}, 2923 (1996)
[arXiv:hep-th/9601135].
}

\lref\HerzogPC{ C.~P.~Herzog and D.~T.~Son, ``Schwinger-Keldysh
propagators from AdS/CFT correspondence,'' JHEP {\bf 0303}, 046
(2003) [arXiv:hep-th/0212072].
}

\lref\SonSD{ D.~T.~Son and A.~O.~Starinets, ``Minkowski-space
correlators in AdS/CFT correspondence: Recipe and applications,''
JHEP {\bf 0209}, 042 (2002) [arXiv:hep-th/0205051].
}

\lref\KlebanovTB{
  I.~R.~Klebanov and E.~Witten,
  ``AdS/CFT correspondence and symmetry breaking,''
  Nucl.\ Phys.\ B {\bf 556}, 89 (1999)
  [arXiv:hep-th/9905104].
}

\lref\MarolfFY{
  D.~Marolf,
  ``States and boundary terms: Subtleties of Lorentzian AdS/CFT,''
  arXiv:hep-th/0412032.
}

\lref\BanksDD{
  T.~Banks, M.~R.~Douglas, G.~T.~Horowitz and E.~J.~Martinec,
  ``AdS dynamics from conformal field theory,''
  arXiv:hep-th/9808016.
}

\lref\BalasubramanianSN{
  V.~Balasubramanian, P.~Kraus and A.~E.~Lawrence,
  ``Bulk vs. boundary dynamics in anti-de Sitter spacetime,''
  Phys.\ Rev.\ D {\bf 59}, 046003 (1999)
  [arXiv:hep-th/9805171]; \quad
  V.~Balasubramanian, P.~Kraus, A.~E.~Lawrence and S.~P.~Trivedi,
  ``Holographic probes of anti-de Sitter space-times,''
  Phys.\ Rev.\ D {\bf 59}, 104021 (1999)
  [arXiv:hep-th/9808017].
}

\lref\hawkingpage{
  S.~W.~Hawking and D.~N.~Page,
  ``Thermodynamics Of Black Holes In Anti-De Sitter Space,''
  Commun.\ Math.\ Phys.\  {\bf 87}, 577 (1983).
}



\lref\HorowitzPQ{
  G.~T.~Horowitz and S.~F.~Ross,
  ``Possible resolution of black hole singularities from large N gauge
  theory,''
  JHEP {\bf 9804}, 015 (1998)
  [arXiv:hep-th/9803085].
}


\lref\BalasubramanianZV{ V.~Balasubramanian and S.~F.~Ross,
``Holographic particle detection,'' Phys.\ Rev.\ D {\bf 61},
044007 (2000) [arXiv:hep-th/9906226].
}

\lref\HemmingKD{
  S.~Hemming, E.~Keski-Vakkuri and P.~Kraus,
  ``Strings in the extended BTZ spacetime,''
  JHEP {\bf 0210}, 006 (2002)
  [arXiv:hep-th/0208003].
}

\lref\KaplanQE{
  J.~Kaplan,
  ``Extracting data from behind horizons with the AdS/CFT correspondence,''
  arXiv:hep-th/0402066.
}

\lref\LoukoTP{ J.~Louko, D.~Marolf and S.~F.~Ross, ``On geodesic
propagators and black hole holography,'' Phys.\ Rev.\ D {\bf 62},
044041 (2000) [arXiv:hep-th/0002111].
}

\lref\LeviCX{ T.~S.~Levi and S.~F.~Ross, ``Holography beyond the
horizon and cosmic censorship,'' arXiv:hep-th/0304150.
}

\lref\GregoryAN{ J.~P.~Gregory and S.~F.~Ross, ``Looking for event
horizons using UV/IR relations,'' Phys.\ Rev.\ D {\bf 63}, 104023
(2001) [arXiv:hep-th/0012135].
}

\lref\BalasubramanianZU{ V.~Balasubramanian and T.~S.~Levi,
``Beyond the veil: Inner horizon instability and holography,''
arXiv:hep-th/0405048.
}

\lref\BrecherGN{
  D.~Brecher, J.~He and M.~Rozali,
  ``On charged black holes in anti-de Sitter space,''
  arXiv:hep-th/0410214.
}

\lref\starinets{ A.~Nunez and A.~O.~Starinets, ``AdS/CFT
correspondence, quasinormal modes, and thermal correlators in  N =
4 SYM,'' arXiv:hep-th/0302026.
}

\lref\kos{ P.~Kraus, H.~Ooguri and S.~Shenker, ``Inside the
horizon with AdS/CFT,'' arXiv:hep-th/0212277.
}

\lref\gkp{ S.~S.~Gubser, I.~R.~Klebanov and A.~M.~Polyakov,
``Gauge theory correlators from non-critical string theory,''
Phys.\ Lett.\ B {\bf 428}, 105 (1998) [arXiv:hep-th/9802109].
}

\lref\MaldacenaRE{ J.~M.~Maldacena, ``The large N limit of
superconformal field theories and supergravity,'' Adv.\ Theor.\
Math.\ Phys.\  {\bf 2}, 231 (1998) [Int.\ J.\ Theor.\ Phys.\  {\bf
38}, 1113 (1999)] [arXiv:hep-th/9711200].
}

\lref\witten{ E.~Witten, ``Anti-de Sitter space and holography,''
Adv.\ Theor.\ Math.\ Phys.\  {\bf 2}, 253 (1998)
[arXiv:hep-th/9802150].
}

\lref\maldat{ J.~M.~Maldacena, ``Eternal black holes in
Anti-de-Sitter,'' arXiv:hep-th/0106112.
}

\lref\witt{ E.~Witten, ``Anti-de Sitter space, thermal phase
transition, and confinement in gauge theories,'' Adv.\ Theor.\
Math.\ Phys.\  {\bf 2}, 505 (1998) [arXiv:hep-th/9803131].
}

\lref\sussWi{
  L.~Susskind and E.~Witten,
  ``The holographic bound in anti-de Sitter space,''
  arXiv:hep-th/9805114.
}

\lref\shenker{
  L.~Fidkowski, V.~Hubeny, M.~Kleban and S.~Shenker,
  ``The black hole singularity in AdS/CFT,''
  JHEP {\bf 0402}, 014 (2004)
  [arXiv:hep-th/0306170].
}

\lref\mot{
  L.~Motl and A.~Neitzke,
  ``Asymptotic black hole quasinormal frequencies,''
  Adv.\ Theor.\ Math.\ Phys.\  {\bf 7}, 307 (2003)
  [arXiv:hep-th/0301173].
}

\lref\ricar{
  V.~Cardoso, J.~Natario and R.~Schiappa,
  ``Asymptotic quasinormal frequencies for black holes in non-asymptotically
  flat spacetimes,''
  J.\ Math.\ Phys.\  {\bf 45}, 4698 (2004)
  [arXiv:hep-th/0403132].
}

\lref\FLlong{G.~Festuccia and H.~Liu, to appear.}

\lref\FLii{G.~Festuccia and H.~Liu, to appear.}

\lref\FLSup{G.~Festuccia, H.~Liu and A.~Scardicchio, unpublished.}

\lref\horomal{
  G.~T.~Horowitz and J.~Maldacena,
  ``The black hole final state,''
  JHEP {\bf 0402}, 008 (2004)
  [arXiv:hep-th/0310281].
}

\lref\bekenstein{
  J.~D.~Bekenstein and L.~Parker,
  ``Path Integral Evaluation Of Feynman Propagator In Curved Space-Time,''
  Phys.\ Rev.\ D {\bf 23}, 2850 (1981).
}

\lref\HertogRZ{
  T.~Hertog and G.~T.~Horowitz,
  ``Towards a big crunch dual,''
  JHEP {\bf 0407}, 073 (2004)
  [arXiv:hep-th/0406134].
}

\lref\HertogHU{
  T.~Hertog and G.~T.~Horowitz,
  ``Holographic description of AdS cosmologies,''
  arXiv:hep-th/0503071.
}


\lref\siopsis{
  G.~Siopsis,
  ``Large mass expansion of quasi-normal modes in AdS(5),''
  Phys.\ Lett.\ B {\bf 590}, 105 (2004)
  [arXiv:hep-th/0402083].
}

\lref\berryG{
  M.~V.~Berry,
  ``Infinitely many Stokes smoothings in the gamma
 function,'' Proc. \ Roy. \ Soc. \  Lond. \
A {\bf 434}, 465 (1991).}

\lref\loOr{
  See e.g. G.~T.~Horowitz and A.~R.~Steif,
  ``Singular string solutions with nonsingular initial data,''
  Phys.\ Lett.\ B {\bf 258}, 91 (1991).
}

\lref\HamiltonJU{
  A.~Hamilton, D.~Kabat, G.~Lifschytz and D.~A.~Lowe,
  ``Local bulk operators in AdS/CFT: A boundary view of horizons and
  locality,''
  arXiv:hep-th/0506118.
}

\lref\FidkowskiFC{
  L.~Fidkowski and S.~Shenker,
  ``D-brane instability as a large N phase transition,''
  arXiv:hep-th/0406086.
}

\lref\DanielssonZT{
  U.~H.~Danielsson, E.~Keski-Vakkuri and M.~Kruczenski,
  ``Spherically collapsing matter in AdS, holography, and shellons,''
  Nucl.\ Phys.\ B {\bf 563}, 279 (1999)
  [arXiv:hep-th/9905227].
}

\lref\DanielssonFA{
  U.~H.~Danielsson, E.~Keski-Vakkuri and M.~Kruczenski,
  ``Black hole formation in AdS and thermalization on the boundary,''
  JHEP {\bf 0002}, 039 (2000)
  [arXiv:hep-th/9912209].
}


\lref\NatarioJD{
  J.~Natario and R.~Schiappa,
  ``On the classification of asymptotic quasinormal frequencies for
  d-dimensional black holes and quantum gravity,''
  arXiv:hep-th/0411267.
}



\Title{\vbox{\baselineskip12pt \hbox{hep-th/0506202}
\hbox{MIT-CTP-3641}
}}%
 {\vbox{\centerline{Excursions beyond the horizon:}
 \medskip
 \smallskip
 {Black hole singularities in
 Yang-Mills
 theories (I) }} }

\smallskip
\centerline{Guido Festuccia and Hong Liu }
\medskip

\centerline{\it  Center for Theoretical Physics} \centerline{\it
Massachusetts Institute of Technology} \centerline{\it Cambridge,
Massachusetts, 02139}

\smallskip

\smallskip

\smallskip

\vglue .3cm

\bigskip
\noindent

We study  black hole singularities in the AdS/CFT correspondence.
These singularities show up in CFT in the behavior of
finite-temperature correlation functions. We first establish a
direct relation between space-like geodesics in the bulk and
momentum space Wightman functions of CFT operators of large
dimensions. This allows us to probe the regions inside the horizon
and near the singularity using the CFT. Information about the
black hole singularity is encoded in the exponential falloff of
finite-temperature correlators at large imaginary frequency. We
construct new gauge invariant observables whose divergences
reflect the presence of the singularity. We also find a UV/UV
connection that governs physics inside the horizon. Additionally,
we comment on the possible resolution of the singularity.

 \Date{June 22, 2005}


\bigskip



\newsec{Introduction}

The AdS/CFT correspondence~\refs{\MaldacenaRE,\witten,\gkp}
provides exciting avenues for exploring various issues in quantum
gravity. An important question that the AdS/CFT correspondence may
shed light on is that of the nature of spacelike singularities,
like the Big Bang or the singularity of a Schwarzschild black
hole. A good laboratory for studying spacelike singularities is an
eternal black hole in anti-de Sitter (AdS) spacetime. It has been
conjectured that quantum gravity in an AdS$_{d+1}$ black hole
background is described by a boundary conformal field theory on
$S^{d-1} \times \IR$ at a temperature given by the Hawking
temperature of the black
hole~\refs{\witten,\witt,\maldat}\foot{More precisely, an AdS
black hole appears as a saddle point in the path integral of
boundary theories in the large $N$ limit.}.

The conventional wisdom regarding singularities is that they
signal the breakdown of classical gravity and should go away when
stringy or quantum gravitational effects are taken into account.
Since in AdS/CFT, classical gravity corresponds to the large $N$
and large 't Hooft coupling limit of the boundary
theory\foot{While we are interested in black holes in all
dimensions, when talking about boundary theories, we will more
specifically have in mind $d=4$, which are $SU(N)$ Yang-Mills
theories.}, one expects that finite $N$ or finite 't Hooft
coupling effects may resolve these
singularities~\refs{\HorowitzPQ,\maldat}. Such considerations
suggest the following strategy:

\item{1.}  Identify manifestations of the black hole singularity
in the large $N$ and large 't Hooft coupling limit of the finite
temperature boundary theory;

\item{2.} From these manifestations, understand the precise
physical mechanism through which the finite $N$ or finite 't Hooft
coupling effects may resolve the singularity.

\ndt  In the boundary theory, the physical observables are
correlation functions of gauge invariant operators. This means
that the physics of singularities should be encoded in the
behavior of boundary correlation functions in appropriate limits.

One of the obstacles\foot{in addition to the standard difficulties
to decode the local bulk physics from boundary theories} in
understanding black hole singularities from finite temperature
boundary theory is that the singularities are hidden behind event
horizons. The boundary conformal field theory evolves through the
bulk Schwarzschild time, i.e. from the point of view of an
external observer, and does not appear to directly describe the
physics beyond the horizon. In other words, if time evolution
inside the horizon of a black hole is to be described by the
boundary theory, time has to be {\it holographically} generated.
This makes the problem particularly challenging, while at the same
time exciting. In particular, understanding physics beyond the
horizon should shed light on how to holographically describe a Big
Bang cosmological spacetime\foot{The regions around the past and
future singularities of a black hole can also be viewed as Big
Bang or Big Crunch cosmologies.}.

A number of
authors~\refs{\BalasubramanianZV\LoukoTP\maldat\kos\LeviCX\HemmingKD\shenker\BalasubramanianZU\FidkowskiFC\KaplanQE\BrecherGN-\HamiltonJU
} have explored how to extract physics beyond the horizon from the
boundary theory correlation functions\foot{See
also~\refs{\DanielssonFA,\DanielssonZT}. For other recent
discussion of spacelike singularities in AdS/CFT, see e.g.
~\refs{\horomal,\HertogHU,\HertogRZ}.}. In particular, Fidkowski
et al~\refs{\shenker} found an interesting but subtle signal of
the singularity in the boundary correlators. They found that
AdS$_{d+1}$ black holes with dimension $d \geq 3$ contain
spacelike geodesics, connecting two asymptotic boundaries, which
could get arbitrarily close to the singularity. The authors
further argued that such geodesics imply the presence of poles on
secondary sheets of the analytically continued coordinate space
correlation functions in the large operator dimension limit.

Here we further explore the manifestations of the black hole
singularity in the boundary theory and discuss their implications
for the resolution of the singularity.

We will establish a direct relation between space-like geodesics
in the bulk spacetime and the large operator dimension limit of
momentum space Wightman functions in the boundary theory. We show
that  physics in the region beyond the horizon is encoded in the
behaviors of boundary correlation functions along the imaginary
frequency axis. In particular, this gives a clear indication that
the ``time'' inside the horizon is holographically generated from
the boundary theory. The presence of the curvature singularity
leads to certain exponential falloff of the correlation functions
near the imaginary infinity. We also construct new gauge invariant
observables which have singularities precisely reflecting the
curvature singularity of the black hole.

In this paper we will present the main idea using the example of
an AdS$_5$ Schwarzschild black hole, leaving technical details and
more extensive discussions of other examples to a longer companion
paper~\refs{\FLlong}. In~\refs{\FLii} we develop a new method for
computing the large mass quasi-normal frequencies of the black
hole, whose knowledge is crucial for the discussion of this paper
and~\refs{\FLlong}.

The plan of the paper is as follows. In section 2, we review the
relevant black hole geometry and the computation of boundary
Wightman function in AdS/CFT. In section 3, we establish a
connection between bulk spacelike geodesics and boundary Wightman
functions in the large operator dimension limit. In section 4, we
study the manifestation of the singularities in Yang-Mills theory.
We conclude in section 5 with a discussion of the possible
resolution of singularities at finite $N$.

\newsec{Boundary correlation functions from AdS/CFT }

\subsec{Black hole geometry}

We will consider big black holes in AdS$_5$, which have a positive
specific heat and are the dominant contribution to the thermal
canonical ensemble of the boundary Yang-Mills theory when the
temperature is sufficiently high~\refs{\hawkingpage}.

The metric for a Schwarzschild black hole in an AdS$_{5}$
spacetime is given by
 \eqn\adsbh{
ds^2=-f(r)dt^2+f(r)^{-1}dr^2+r^2 d\Omega_{3}^2
 }
with
 \eqn\defF{
f(r)=r^2+1-{\mu \ov r^{2}} = {1 \ov
r^2}(r^2-r_0^2)(r^2+r_1^2),\qquad
 }
 $$
r_1^2=1+r_0^2, \qquad \mu = r_0^2 r_1^2
 $$
where $\mu$ is proportional to the mass of the black hole and the
event horizon is at $r=r_0$. $r_0, r_1$ can be solved in terms of
$\mu$. We have set the curvature radius of AdS to be unity, as we
will do throughout the paper. As $r \to \infty$, the metric goes
over to that of global AdS with $t$ identified as the boundary
time. The fully extended black hole spacetime has two disconnected
time-like boundaries, each of topology $S^3 \times \IR$.

\ifig\penrose{Penrose diagram for the AdS black hole. There are
two asymptotic AdS regions, which are space-separated from each
other. A null geodesic going from the boundary to the singularity
is indicated in the figure.} {\epsfxsize=4cm
\epsfbox{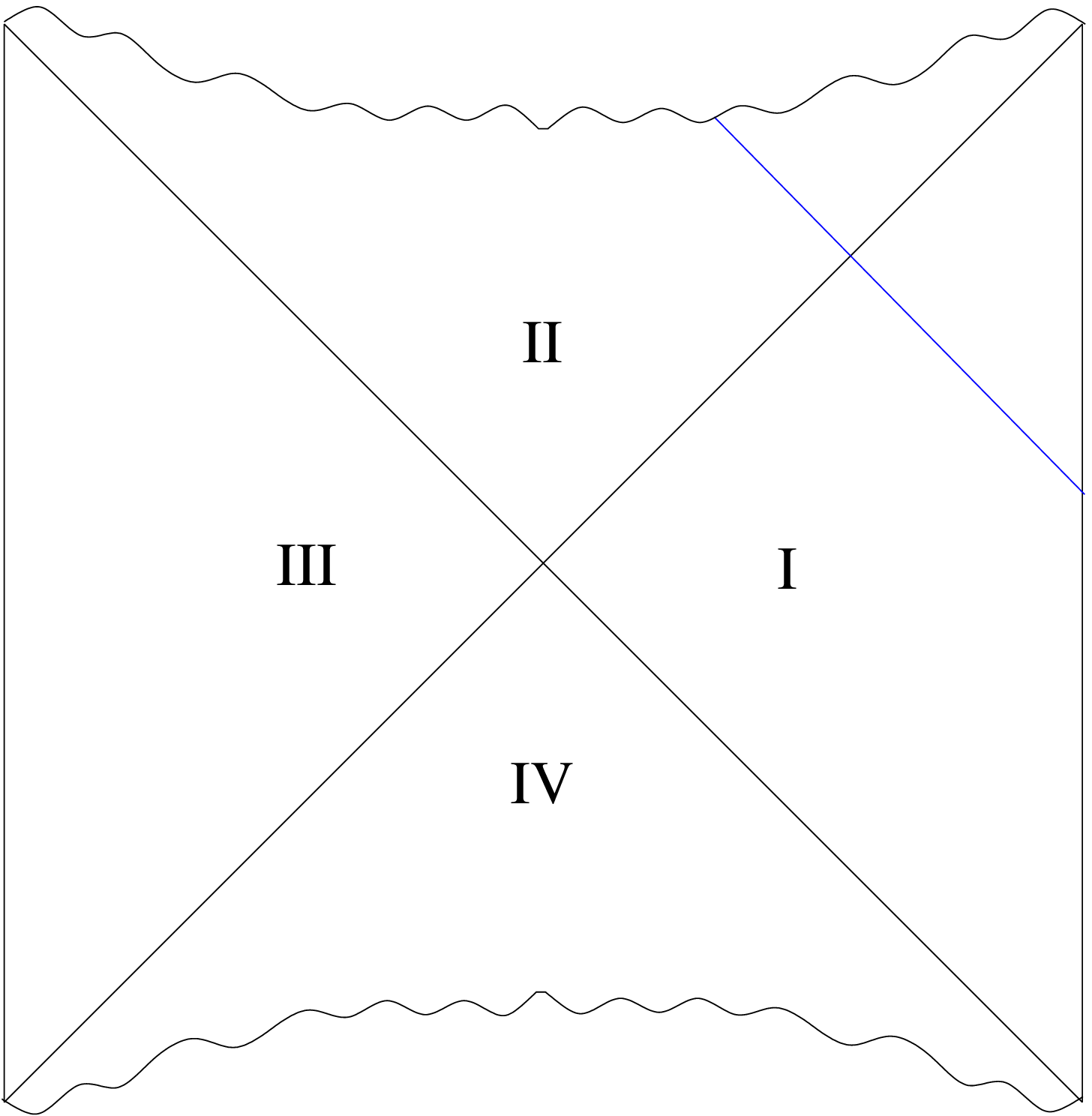}}

It is often convenient to use the tortoise coordinate
 \eqn\tortoise{
 z =  \int_{r}^\infty {dr \ov f(r)} = -{\beta \ov 4 \pi} \log \le({r-r_0 \ov r+r_0}\ri) + {\tilde
\beta \ov 2 \pi} \tan^{-1} {r_1 \ov r}
 }
with
 \eqn\hawkGf{ \beta = {2 \pi r_0 \ov r_0^2 + r_1^2}, \qquad
\tilde \beta =  {2 \pi r_1 \ov r_0^2 + r_1^2} \ .
 }
The region outside the horizon corresponds to $z \in (0, +\infty)$
with $z=0$ at the boundary and $z \to + \infty$ at the horizon.
$\beta$ is the inverse Hawking temperature. To have a feeling of
the physical meaning of $\tilde \beta$ we note that the {\it
complex Schwarzschild time} that it takes for a radial null
geodesic to go from the boundary to the singularity is given by
 \eqn\geTi{
   \pm \int_0^\infty {dr \ov f(r)} = \pm  {1 \ov 4} (\tilde \beta \pm i \beta)
 }
where the imaginary part of the integral arises by going around
the pole at $r=r_0$ in the complex $r$-plane. A nonzero $\tilde
\beta$ implies that the Penrose diagram for the black hole is not
a square, as was first pointed out in~\refs{\shenker}. It is
convenient to introduce a complex quantity
 \eqn\defBB{
 \cB   =  \tilde \beta + i \beta  = {2 \pi (r_1 + i r_0) \ov r_0^2 + r_1^2}
  = {2 \pi \ov r_1 - i r_0} \
 }
which will be important in our discussion. One can invert
\tortoise\ to find $r(z)$. In particular, for ${\rm Re} z >
{\tilde \beta \ov 4}$, $r$ is a one-to-one periodic function of
$z$ with period $i {\beta \ov 2}$.

We will consider the complexified Kruskal spacetime in which
points related by
 \eqn\idenGd{
 t \to t + i {m+n \ov 2} \beta , \qquad z \to z + i {m-n \ov 2}
 \beta, \quad m,n \in \IZ
 }
are identified. The Lorentzian section (\penrose) of the
complexified spacetime can be conveniently described using $(t,z)$
with constant imaginary parts. For example, up to identifications
\idenGd, region III can be specified by
 \eqn\imgas{
 {\rm Im} \; t = -{i \beta \ov 2} , \quad  {\rm Im} \; z = 0 \ .
 }

\subsec{Boundary Wightman functions}

Now consider an operator $\OO$ in the boundary theory
corresponding to a bulk field $\phi$ of mass $m$. In the
supergravity limit, the conformal dimension of $\OO$ is given
by~\refs{\gkp,\witten}
 \eqn\defDim{
 \Delta = {d \ov 2} + \nu, \qquad \nu = \sqrt{{d^2 \ov 4} + m^2} \
 .
 }
Thermal boundary two-point functions of $\OO$ can be obtained from
free bulk Green functions of $\phi$ in the Hartle-Hawking vacuum
by taking the arguments of $\phi$ to the boundary (see
e.g.~\refs{\BanksDD,\KlebanovTB}).\foot{For discussion of boundary
Lorentzian correlation functions in the supergravity
approximation, see also~\refs{\BalasubramanianSN,
\SonSD,\HerzogPC,\MarolfFY}.} For example, the boundary Wightman
function is obtained by
 \eqn\bDCo{\eqalign{
 G_+ (x,x') & = \lim_{r,r' \to \infty} (2
 \nu r^{\Delta} ) (2 \nu r'^{\Delta} )  \GG_+ (x,r; x',r') \cr
 }}
 where $\GG_+$ and $G_+$ denote the bulk and boundary
correlation functions respectively
 \eqn\Gont{\eqalign{
 G_{+} (x,x') = \Tr \le[e^{-\beta H} \OO (x) \OO (x')
 \ri],
 }}
 \eqn\Gontb{
  \GG_+ (x,r; x',r') = \vev{0|\phi (r,x)
 \phi(r',x')|0}_{HH} \ .
 }
In the above equations we used the notation $x= (t,  e)$ with $e$
denoting a point on $S^3$ and the subscript ``$HH$'' denotes the
Hartle-Hawking vacuum. Going to momentum space \bDCo\ becomes (see
Appendix A for a definition of the Fourier transform)
 \eqn\MbDCo{\eqalign{
 G_+ (\om, l) & = \lim_{r,r' \to \infty} (2
 \nu r^{\Delta} ) (2 \nu r'^{\Delta} )  \GG_+ (\om, l;r,r') \ , \cr
 }}
where $l$ denotes the angular momentum on $S^3$. The Feynmann
(retarded) propagator
 in the bulk leads to the
Feynmann (retarded) Green function on the boundary by the same
procedure. In this paper we will focus on $G_+$ for reasons to be
commented on later.

Since the extended black hole background has two asymptotic
boundaries, we can also take $r$ and $r'$ to different boundaries.
Such points are always space-like separated in the bulk and  lead
to boundary correlation functions of complex time separation (see
equation \imgas)
 \eqn\Gont{\eqalign{
 G_{12} (t) = \Tr \le[e^{-\beta H} \OO (t-i\beta/2) \OO (0)
 \ri]
 = G_+ (t-i \beta/2)  \  \cr
 }}
where we have suppressed the boundary spatial coordinates for
notational simplicity. $G_{+} (t)$ is analytic for
 $-{\beta } < {\rm Im} t < 0 $ and the
two-sided correlator $G_{12} (t) $ can be obtained from $G_+ (t)$
by simple analytic continuation\foot{In contrast, the Feynman and
retarded functions have singularities in the range $-{\beta \ov 2}
< {\rm Im} t < 0$.}.

$\GG_+$ can be found in terms of solutions to the Laplace equation
for $\phi$ following the standard free field quantization
procedure. Let
$$
\phi = e^{- i \om t} Y_I (e) r^{-{d-1 \ov 2}} \psi (\om, p;r),
$$
with $Y_I (e)$ denoting scalar spherical harmonics on $S^3$ (see
appendix A for notations on spherical harmonics). The Laplace
equation for $\phi$ can then be written in terms of the tortoise
coordinate as a Schrodinger equation for $\psi$
 \eqn\TeomD{ \eqalign{
  \left(-  \p_z^2   + V_l (z) - \om^2 \right) \psi =0 \ .
  }}
$V_l (z) $ can be expressed through $r(z)$ as
 \eqn\poeV{
 V_l(z) = f(r) \left[ {(l+1)^2-{1 \ov 4} \ov r^2} +\nu^2 -{1 \ov 4}
 + {9 \mu \ov 4 r^4} \right] \
 }
For real $l>0$, $V_l (z)$ is a monotonically decreasing function
of $z \in (0, \infty)$. Near the boundary,
 \eqn\bougd{
V_l \approx {\nu^2 - {1 \ov 4} \ov z^2}, \qquad z \to 0 \ ,
 }
and near the horizon
 \eqn\fallho{
V_l  \propto e^{-{4 \pi \ov \beta} z } \to 0 , \qquad z \to
+\infty \ .
 }

For any given real $\om$ the Schrodinger equation \TeomD\ has a
unique normalizable mode $\psi_{\om l}$, which we will take to be
real. We normalize it at the horizon as ($\delta_\om$ is a phase
shift)
 \eqn\horiB{ \psi_{\om
 l} (z) \approx  e^{- i \om z - i \delta_{\om}} + e^{i \om z + i
 \delta_{\om}}, \qquad z \to + \infty \ .
 }
As $z \to 0$, $\psi_{\om l}$ has the form
 \eqn\solIn{
\psi_{\om l} \approx C(\om,l) z^{\ha +\nu} + \cdots, \qquad z \to
0
 }
where the constant $C$ is fixed by the normalization of \horiB. It
is easy to check that $\psi_{\om l}$ is even in $\om$.

Using the mode expansion of $\phi$ in the Hartle-Hawking vacuum,
the bulk Wightman propagator $\GG_+$ \Gont\ in momentum space can
be written in terms of $\psi_{\om l}$ as
 \eqn\bulkP{\eqalign{ \GG_+ (\om,
l;r,r') & =  {1 \ov 2 \om} {e^{\beta \om} \ov e^{\beta \om} -1} \,
(r r')^{-{d-1 \ov 2}}
 \psi_{\om l} (r) \psi_{\om l} (r') \cr
 }}
which leads to the boundary $G_+$ upon using \MbDCo\ and \solIn
 \eqn\wihE{\eqalign{
 G_+ (\om,l) & = {(2 \nu)^2 \ov 2 \om} {e^{\beta \om} \ov e^{\beta
\om} -1} C^2 (\om,l) \cr
 }}
$G_+$ is to be evaluated for real $\om,  l$ and can be
analytically continued to general complex $\om$ and $l$.

\subsec{Analytic properties}

Equations \bulkP\ and \wihE\ indicate that the boundary Yang-Mills
theory has a {\it continuous} spectrum in the large $N$ and large
't Hooft coupling limit, despite being on a compact space. At
finite $N$, the theory should have a discrete spectrum on $S^3$.
In the bulk the continuous spectrum arises due to the presence of
the horizon.

The analytic properties of $G_+$ in the complex $\om$-plane for a
given $l$ can be deduced by applying standard techniques of
scattering theory to \TeomD. The fact that $r$ is a periodic
function of the tortoise coordinate with a period $i {\beta \ov
2}$ implies that:~\refs{\FLlong}

\item{1.} The poles in the prefactor ${1 \ov  \om (e^{\beta \om}
-1)}$ cancel with zeros of $C^2$. $G_+$ is analytic at $\om=0$ and
$\om = {2 \pi i n \ov \beta}, \;\; n \in \IZ$.

\item{2.}  The only singularities of $G_+$ in the complex
$\om$-plane are poles. The locations of the poles obey a
reflection symmetry: if there is a pole at $\om_0$, then there are
poles at $-\om_0, \om^*_0, - \om^*_0$.

\ndt These two features are quite generic, applicable to AdS black
holes of all dimensions. The poles of $G_+$ in the lower half
$\om$-plane coincide with those of the retarded propagator $G_R$
and correspond to quasi-normal frequencies of the black hole
background. Since it is not known how to solve the Schrodinger
equation \TeomD\ exactly, the determination of the quasi-normal
poles is a difficult mathematical problem.

\ifig\poles{Poles for $G_+ (\om,l)$ for $l=0$ in the complex
$\om$-plane. We use $r_0 =1, r_1 =\sqrt{2}$.} {\epsfxsize=6cm
\epsfbox{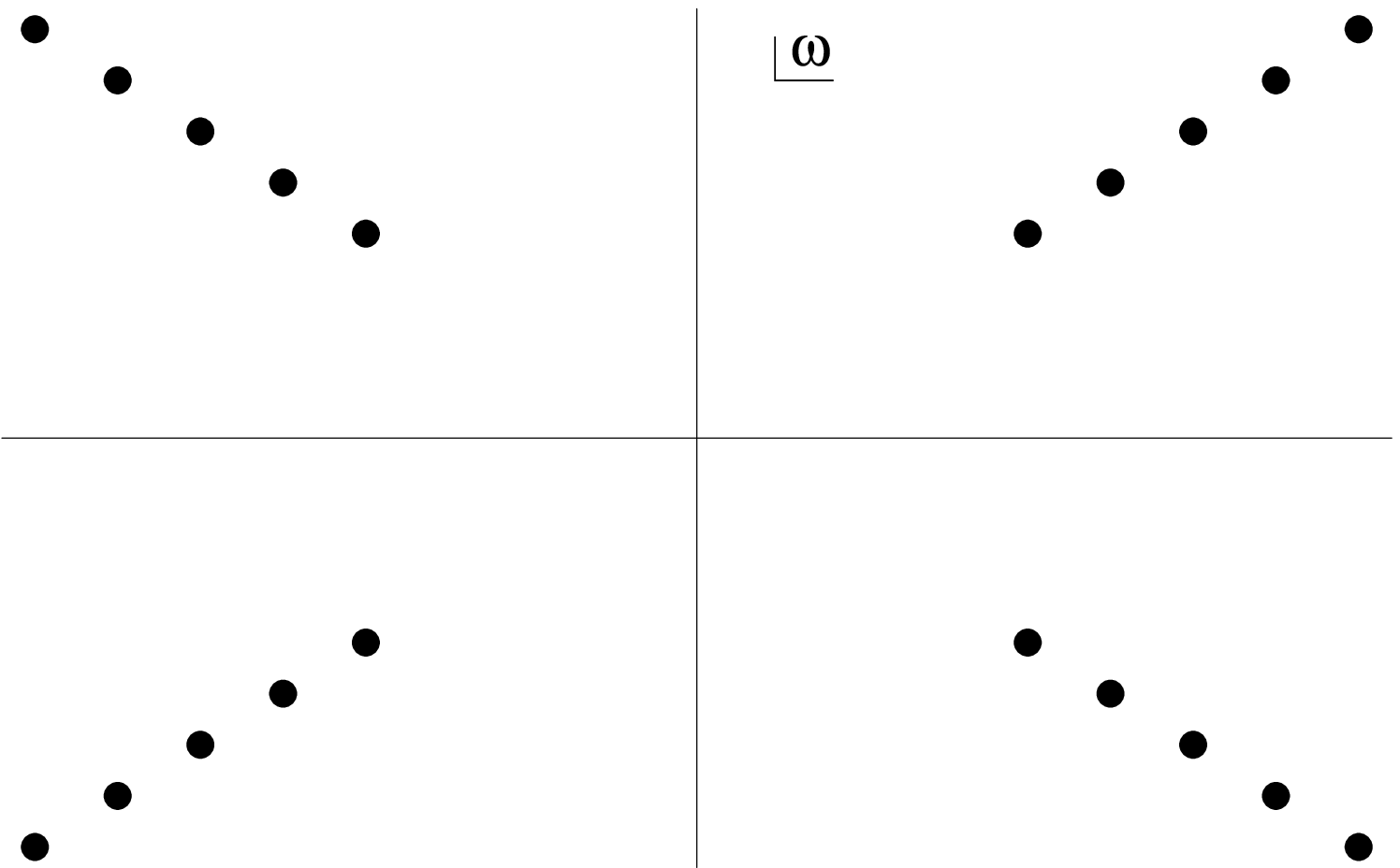}}

When $l=0$ (or small compared to $\nu$ or $\om$), the problem
simplifies and various
methods~\refs{\starinets,\mot,\ricar,\NatarioJD,\siopsis} can be
used to determine the locations of poles of $G_+$ approximately.
One finds that there are four infinite lines of poles as indicated
in \poles. The poles in the upper right quadrant are given
by~\refs{\FLii}
 \eqn\locaP{
 \om \approx { 2 \pi \ov \bar \cB} \nu + \om_0 +  { 4 \pi n \ov \bar \cB}
 , \qquad
  \quad n=0,1, \cdots
 }
The other lines are obtained by reflections. $\om_0$ in \locaP\ is
a constant of $O(1)$ (independent of $\nu$) whose exact value is
not relevant here\foot{The accuracy of equation \locaP\ increases
with $n$. A comparison with the numerical results obtained
in~\refs{\starinets} shows that it appears to work well even for
$n$ small.}.

 The quasi-normal frequencies for $l \neq 0$ and other dimensions are more
complicated to find and will be discussed in~\refs{\FLii}.

\newsec{Boundary Wightman functions and bulk geodesics}

\subsec{A semi-classical approximation}

We now develop a ``semi-classical'' approximation to equation
\TeomD, in the following large $\nu$ limit
 \eqn\largnu{
 \om = \nu u,  \qquad l+1 = \nu k , \qquad  \nu \gg  1 \ ,
 }
i.e. we take the mass $m$ of $\phi$ to be large and ``measure''
the frequency $\om$ and angular momentum $l$ in units of $m$. With
$\psi = e^{\nu S}$ equation \TeomD\ becomes
 \eqn\liuoG{
 - (\p_z S)^2 - {1 \ov \nu} \p_z^2 S + V(z)  +
 {1 \ov \nu^2} Q(z) = u^2
 }
with
 \eqn\degFV{
 V(z) = f(r) \le(1 + {k^2 \ov r^2} \ri), \qquad
Q(z) = f(r) \left[ -{1 \ov 4 r^2}  -{1 \ov 4}
 + {9 \mu \ov 4 r^4} \right] \ .
 }

 \ifig\wkb{The potential $V(z)$ with $\mu=10$ and $k=0$ is shown. $z_c$ is the turning point.
 Dashed and solid
 lines indicate the classically forbidden and allowed regions respectively.}
 {\epsfxsize=6cm \epsfbox{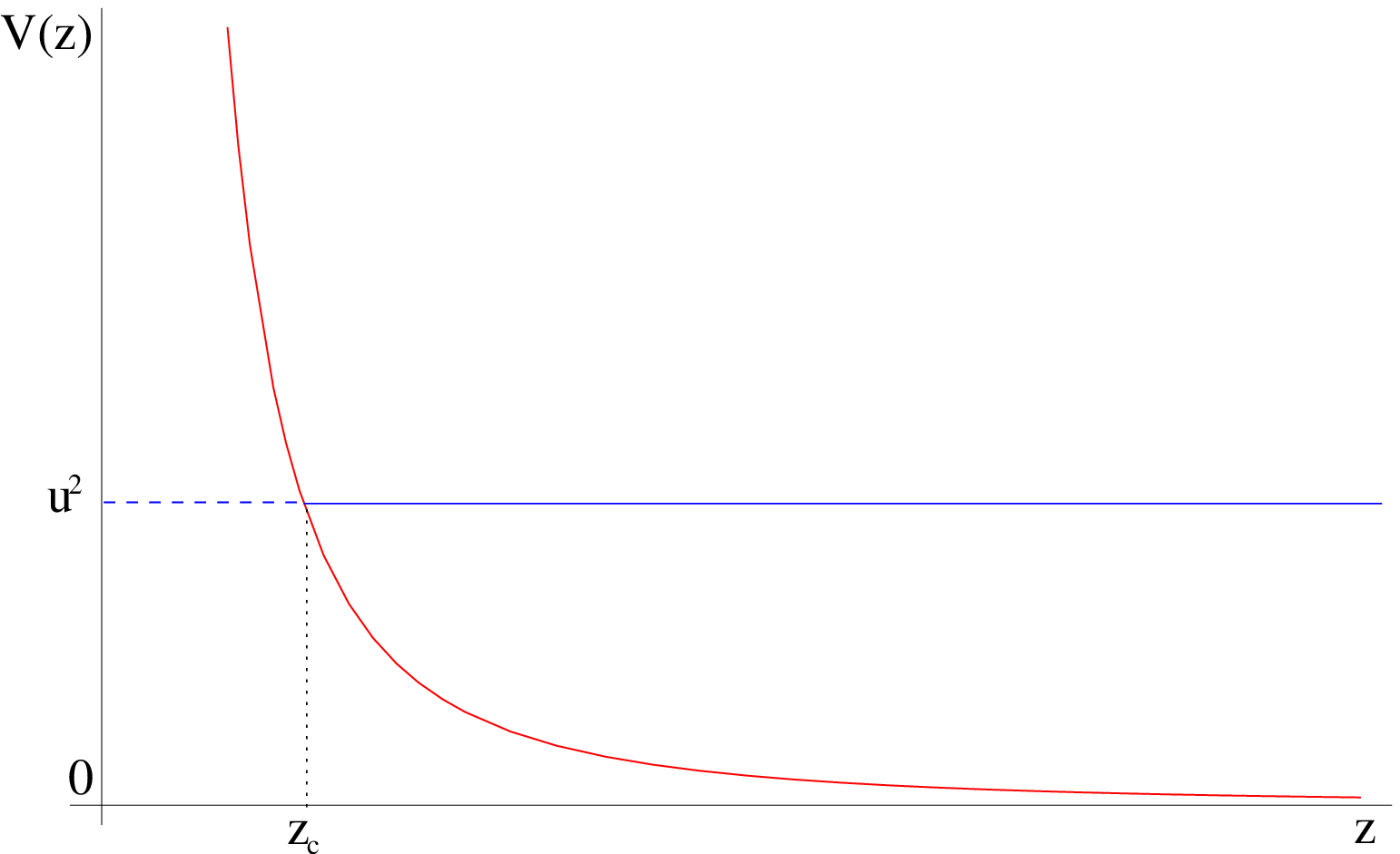}}

With $k^2 \geq 0$ the leading order potential $V(z)$ is a
monotonically decreasing function for $z \in (0, +\infty)$ as
indicated in \wkb.  For scattering sates with $u
>0$, \liuoG\ can be solved order by order in $1/\nu$ expansion
using the standard WKB method. In the classically forbidden region
(see \wkb), the exponentially decreasing solution can be written
as
 \eqn\splg{
 \psi_{\om l}^{(wkb)} (r)
 ={1 \ov \sqrt{f \kappa_r}}e^{ \nu \ZZ
 } \le( 1 + O\le(\nu^{-1} \ri) \ri)
 }
with\foot{The branch cuts for $\kappa_r$ on the complex $r$-plane
are chosen so that they do not intersect the integration contour
in $\ZZ$.}
 \eqn\zzdef{\eqalign{
 \ZZ (r) & = - \int_{r_c}^r dr' \, \kappa_r (r') , \qquad
  \kappa_r  = {1 \ov f} \sqrt{V(r) - u^2} \ .
 }}
$r_c$ in the lower integration limit of \zzdef\ is the turning
point, given by the real positive root of the equation
 \eqn\turnPw{
 V(r) = f (r)\le(1 + {k^2 \ov r^2}\ri) = u^2  \ .
 }
For $u^2>0$, equation \turnPw\ has a unique positive root $ r_c
>r_0$. $\ZZ$ satisfies the equation
 \eqn\wkbeR{\eqalign{
 &  f \ZZ'^2  - {k^2 \ov r^2} + {1 \ov f} u^2 = 1   \  \cr
 }}
with $\ZZ'(r_c) =0$.  Note that we have written the above
equations in terms of $r$ for convenience. One can equivalently
write them in terms of the tortoise coordinate $z$. The
expressions in terms of $r$ are more convenient to visualize the
analytic continuation to be discussed later.

The expression for $\psi^{(wkb)}_{\om l}$ in the classically
allowed region of the potential \degFV\ (i.e. for $z_c < z $ or
$r_0 < r < r_c$) follows from the standard connection formula,
from which one can determine the relative normalization between
$\psi_{\om l}^{(wkb)}$ of \splg\ and $\psi_{\om l}$ of \horiB\ to
be
 \eqn\nwFg{
\psi_{\om l}^{(wkb)}  = {1 \ov \sqrt{u}} \psi_{\om l}, \qquad u >
0, \quad \nu \to \infty
 }

From \nwFg\ we find that in the limit \largnu\ the boundary
Wightman function $G_+$  can be expanded as
 \eqn\bdgpE{
G_+ (\om, l) \approx 2 \nu \; e^{\nu Z (u,k)} \le( 1 +
O({\nu^{-1}}) \ri) + {\rm subdominant \;\; terms}, \qquad \om >0
 }
with
 \eqn\ZdefB{\eqalign{
  Z (u,k) 
  & =  2 \lim_{r \to \infty} \le( \log r -
\int_{r_c}^r dr' \, \kappa_r (r')  \ri)  \ . \cr
 }}

Higher order $1/\nu$ corrections in \bdgpE\ can also be obtained
from \liuoG\ using the standard WKB procedure. In particular, the
term proportional to $Q(z)$ will be important at order $\nu^{-1}$.
There could also be subdominant terms in \bdgpE\ coming from
reflections at other (complex) turning points of $V (r)$.

While equations \bdgpE--\ZdefB\ were obtained for $u>0$ and $k
\geq 0$, they can be analytically continued to the full complex
$u$ and $k$-planes. We will show in section 4 that the analytic
continuation allows us to probe the region beyond the horizon.

\subsec{Relation with geodesics}

We expect $Z(u,k)$ in \ZdefB\ to have a simple interpretation in
terms of bulk geodesics. The reason is that in the large mass
limit, the propagation of bulk field $\phi$ should approximately
follow geodesic paths. Thus we expect a direct relation between
the WKB approximation of the last subsection with the geodesic
approximation. The scaling in \largnu\ simply defines $u$ as the
``velocity'' in $t$ direction and $k$ as the ``angular velocity''
on $S^3$.

Due to translational invariance in $t$ and isometries on $S^3$, a
bulk spacelike geodesic is characterized by the integrals of
motion
 \eqn\velocD{
E =  f {dt \ov ds}, \qquad q = {r^2} {d \th \ov ds}
 }
where $s$ is the proper distance and $\th$ denotes the angular
coordinate along the geodesic motion on $S^3$. We  treat geodesics
which are related by a translation in $t$ and on $S^3$ as
equivalent. The geodesic satisfies the equation
 \eqn\geodE{
 {1 \ov f} \le({dr \ov ds}\ri)^2 + {q^2 \ov r^2} - {1 \ov f} E^2
 =1 \ .
 }
Equation \geodE\ is precisely \wkbeR\ with the
identification\foot{The sign choice for the first expression below
corresponds to having the geodesic moving away from the boundary,
while those for the last two equations are for convenience. The
fact $i$ in relating the boundary ``velocities'' $u,k$ to the
velocities $E,q$ of the bulk geodesic is due to that the geodesic
is spacelike.}
 \eqn\geoRe{
f \ZZ' = {dr \ov ds}, \qquad u= i E , \qquad k = i q \ .
 }
$\kappa_r$ of \zzdef\ can be identified as the proper velocity of
the geodesic along the $r$ direction. Thus $Z (u,k)$ can be
associated with a (complex) spacelike geodesic with constants of
motion $E=-i u$ and $q = -i k$, which starts and ends at $r =
+\infty$.\foot{Note that depending on the values of $E$ and $q$
the starting and end points can be on the same or different
boundaries.} More explicitly, one finds that $Z (u,k)$ can be
written as
  \eqn\defLew{
 Z (u,k) = - E t (E,q) - L (E,q) + q  d (E,q)
 }
 where $L (E,q)$ is the (regularized)
proper distance of the geodesic, $t(E,q)$ is  the time separation
and $d (E,q)$ is the proper distance on $S^3$ between the final
and initial points,
 \eqn\varSInE{\eqalign{
 L (E, q)  & =   2  \lim_{r\to \infty} \le(\int_{r_c}^{r} {dr \ov
 \sqrt{f + E^2  - {f \ov r^2} q^2 }} - \log r \ri)
 \cr
 t (E,  q) &=  2 E \int_{r_c}^\infty {dr \ov f
 \sqrt{f + E^2  - {f\ov r^2 } q^2 }}  \,
 \cr
  d (E,  q) & = 2 q \int_{r_c}^\infty
 {dr \ov r^2 \sqrt{f + E^2 - { f  \ov r^2} q^2}} \ . \cr
 }}
Also note the relation
 \eqn\unre{
 {\p Z \ov \p E} =  -t (E,q) \qquad {\p Z \ov \p q} = d (E,q)\
 }
which shows that $L (t,d)$ and $Z(E,q)$ are related by a Legendre
transform.

Note that $E$ and $q$ do not specify a geodesic uniquely.
\varSInE\ defines a complex geodesic with a choice of root $r_c
(E,q)$ of equation \turnPw\ as the turning point and a contour
from $r_c (E,q)$ to $+\infty$. For the same value of $E,q$, a
different choice of the root or a different contour which cannot
be smoothly deformed into the previous one defines a different
complex geodesic. The identification
\defLew\ with the boundary $Z (u,k)$ selects a specific one among
them.

In the above discussions we have concentrated on the Wightman
functions\foot{Wightman functions already contain all the
information of the theory. For example, Feynmann and retarded
functions can be obtained from them.}. A similar relation with
bulk geodesics can be established for retarded and Feynmann
functions in momentum space, whose story is more complicated since
their dependence on $\nu$ is not uniform. Special care is needed
when $\nu$ is an integer\foot{This can be seen also from zero
temperature correlation functions.}. This makes the large $\nu$
limit more subtle. Even at the leading order, one has to take into
account of an infinite number of classical paths in the WKB
approximation~\refs{\FLSup}. Fortunately, all these additional
complications arise due to the asymptotic behavior of $f \sim r^2$
near the boundary of spacetime and do not seem to give additional
insight into the question of physics beyond the horizon.

\subsec{Coordinate space correlation functions}

We now look at the Fourier transform of $G_+ (\om,l)$ to the
coordinate space correlator (see Appendix A for notations)
 \eqn\GfouB{
 G_{+} (t;e, e') = {1 \ov 4 \pi^2} \sum_{l=0}^\infty
 { 2
(l+1)}  C_l (e \cdot e')
  \int_{-\infty}^\infty {d\om \ov 2 \pi } \,
 e^{- i \om t}  \,
  G_{+}( \om, l) \ .
 }
The two-sided correlator \Gont\ can be obtained from \GfouB\ by
taking $ t \to t - i{\beta \ov 2}$, while the Euclidean correlator
$G_E (\tau;e,e')$ can be obtained by taking $t = - i \tau$ with $0
< \tau < \beta$.

In the large $\nu$ limit \largnu, using \bdgpE\ we can approximate
the sum over $l$ in \GfouB\ by an integral over $k$
 \eqn\mewjg{
 G_{+} (t, \th) \approx  {\nu^3 \ov 8 \pi^3 i \sin \th }
 \int_{-\infty}^\infty  d u  d  k \, k \,  e^{-i \nu u t + i \nu k \th} \,
 2 \nu e^{\nu Z (u, k)}
 }
where $\theta = \cos^{-1} (e \cdot e')$ and we have extended the
integration range for $k$ to $(-\infty, \infty)$ using that $Z$ is
an even function of $k$. \mewjg\ can be evaluated by the method of
steepest descent with the saddle points determined by
 \eqn\unrP{
{\p Z \ov \p u} = i t, \qquad  {\p Z \ov \p k} =- i \theta \ .
 }
 Using
equations \unre\ and \geoRe\ we find that \unrP\ become
 \eqn\gsjka{
 t = t (E,q), \qquad \theta = d (E,q)
 }
i.e. bulk geodesics with end point separation given by $(t,\th)$
appear as saddle points of \mewjg.  Since from \defLew\ the
regularized geodesic distance $L (t,\th)$ and $Z(u,k)$ are related
by a Legendre transformation, one finds that
 \eqn\BDCo{\eqalign{
 G_+ (t,\th) & \approx \sum_i 2 \nu J_i^\ha \le({\nu \ov 2
\pi}\ri)^{2} e^{-\nu L_i} \le(1 +
 O(\nu^{-1}) \ri)
 }}
where $i$ sums over the saddles along the steepest descent
contour. The Jacobian $J$ is due to the Gaussian integration
around the saddle points.

Some remarks:

\item{1.}  $J$ can be interpreted as the density of the geodesics.
It is proportional to the Van Vleck-Morette determinant for the
geodesics. One can check that \BDCo\ agrees precisely (including
the prefactor $J$) with the expression obtained directly from the
geodesic approximation to the coordinate space path integral
 \eqn\gEoDA{\eqalign{
 G (x , r;x',r') & = \sum_{paths} e^{{i
 \ov \hbar} m S}  \ , \cr
 }}
after taking the end points to the boundary.

\item{2.}  In the standard geodesic approximation to \gEoDA, it is
often a subtle question in Lorentzian signature to determine which
geodesics contribute to the sum \BDCo. In our approach, the
steepest descent approximation of the  Fourier transform \GfouB\
gives a {\it precise} prescription for determining the sum.

\item{3.}  Higher order terms in \BDCo\ can be computed
systematically in our approach.

\newsec{Black hole singularites in Yang-Mills theory}

In this section we discuss how information about the black hole
singularity can be extracted from $Z(u,k)$ using its relations
with the bulk geodesics developed in the last section. For
simplicity, we restrict our discussion to $k=0$, in which case the
corresponding bulk geodesic has zero angular momentum. The
discussion for general $k$ is more involved and will appear
in~\refs{\FLlong}. The $k=0$ case already captures many of the
essential elements.

\subsec{Probing the physics beyond the horizon and near the
singularity}

We first consider the analytic continuation of $ Z
(u,k=0)$\foot{In the following, we will simply write it as $Z(u)$
and similarly use $G_+ (\om)$ for $G_+ (\om, l=0)$.} to general
complex $u$. We will see that the analytic continuation allows us
to probe the regions beyond the horizon and near the singularity.

As discussed in section 2.3, when $u>0$, the turning point $r_c
(u)$ in \ZdefB\ lies outside the horizon, i.e. $r_c
> r_0$. The integration contour runs along the positive real
$r$-axis from $r_c$ to $+ \infty$. The analytic continuation of $Z
(u)$ to the full complex $u$-plane involves the following two
aspects:

\item{1.} Analytically continue the turning point $r_c (u)$  from
that for real $u>0$;

\item{2.} Smooth deformation of the integration contour as $r_c$
moves on the complex $r$-plane.

\ndt Both steps have some subtleties, which we now discuss in
detail.

\ifig\branchi{The structure of branch cuts of $Z(u)$ and $r_c(u)$
for $r_0=1, r_1 =2$. At finite $\nu$, the branch cuts become the
pole lines of $G_+ (\om, l=0)$ as in \poles. The asymptotic
regions are labelled by $S_{\pm}$ or $B_{\pm}$, indicating whether
the turning point for the corresponding $u$ approaches the
singularity ($S_{\pm}$) or the boundary ($B_{\pm}$).}
{\epsfxsize=6cm \epsfbox{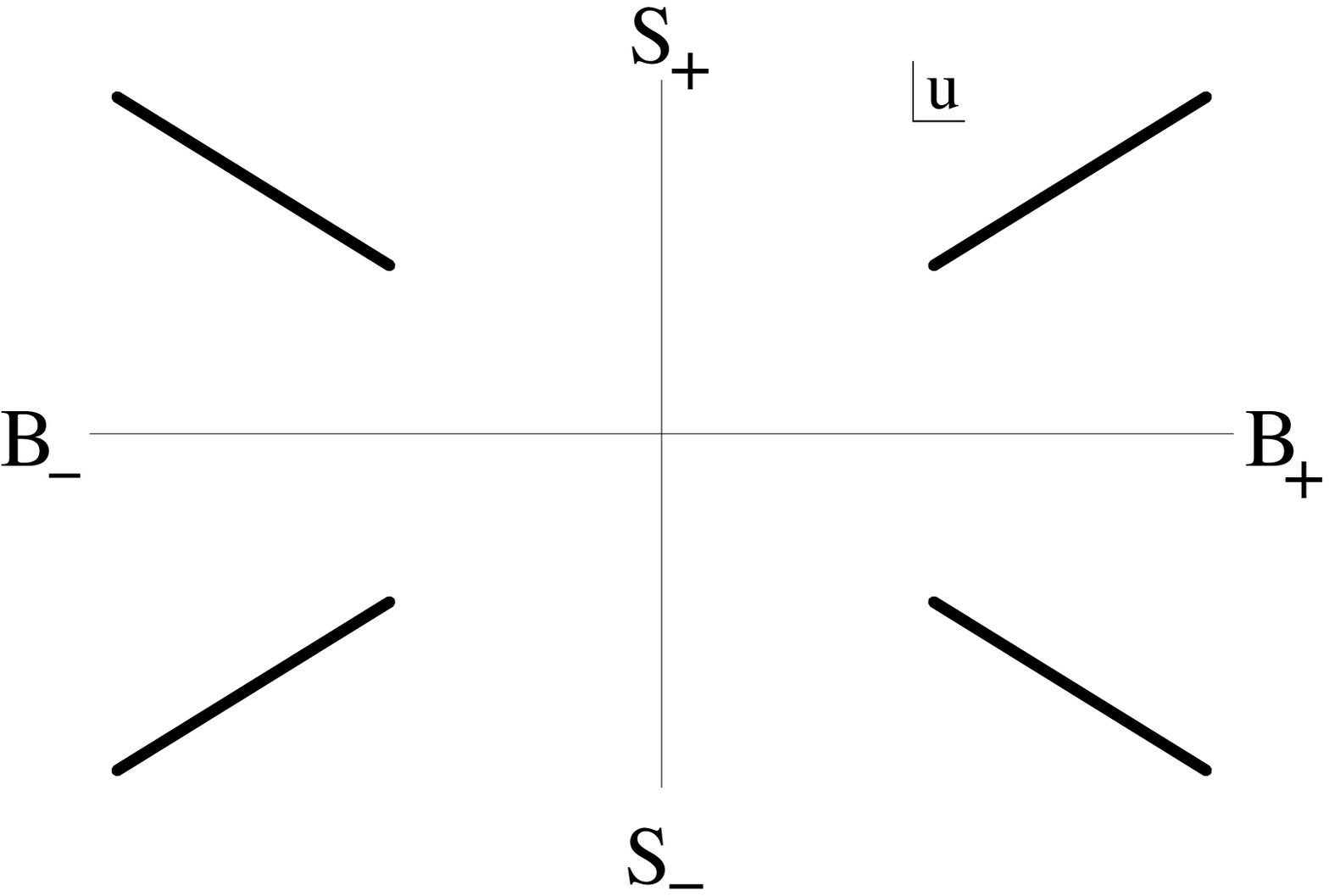}}

The analytic continuation of $r_c$ from the $u>0$ region is not
unique, since $r_c (u)$ has branch points in the complex $u$-plane
at which it merges with other solutions of \turnPw. These are also
branch points of $Z (u)$. When $k=0$ \turnPw\ is a quadratic
equation for $r^2$ and $r_c (u)$ coincides with other roots when
 \eqn\riwa{
 (u^2 -1)^2 + 4 \mu = 0 \ .
 }
From \riwa\ we find that $r_c (u)$ has four branch points at
\eqn\locBr{
 u_0 = \pm (r_1 \pm i r_0) = \pm {2 \pi \ov \bcB}, \pm {2 \pi \ov
 \cB} \ .
 }
For $r_c$ and $Z$ to be single-valued on the $u$-plane, branch
cuts have to be specified. Different choices of the branch cuts
correspond to different ways of performing the analytic
continuation. The locations of the branch cuts cannot be
determined from the integrals \ZdefB\ or \varSInE\ alone. To
determine them we need to use analytic properties of $G_+ (\om)$.
As discussed around equation \locaP, the only singularities of
$G_+ (\om)$ at finite $\nu$ are four lines of poles located at
 \eqn\dhf{
 u \approx  { 2 \pi \ov \bar \cB}+ {\om_0 \ov \nu} +  { 4 \pi n \ov \nu \bar  \cB}
 , \qquad
  \quad n=0,1, \cdots \ .
 }
and  the reflections of \dhf\ with respect to the real and
imaginary $u$ axes. In the large $\nu$ limit, since the spacings
between poles go to zero,
 these lines of poles become branch cuts  of $Z (u)$.\foot{The simplest
 example exhibiting this behavior
 is a Gamma-function $\Ga (\nu z)$. The function has poles along the negative real
 axis for finite $\nu$. In the large $\nu$ limit, upon using the Stirling
formula, the pole line is replaced by a branch cut.}
  This determines the directions
of the branch cuts to be along the radial direction from each
branch point to infinity (see \branchi).

\ifig\Feffpot{A radial spacelike geodesic can be described by a
particle of energy $-u^2$ moving in the potential $U=-f$. The
horizon is at $r=r_0$. For $u^2<0$, the turning point lies inside
the horizon.} {\epsfxsize=6cm \epsfbox{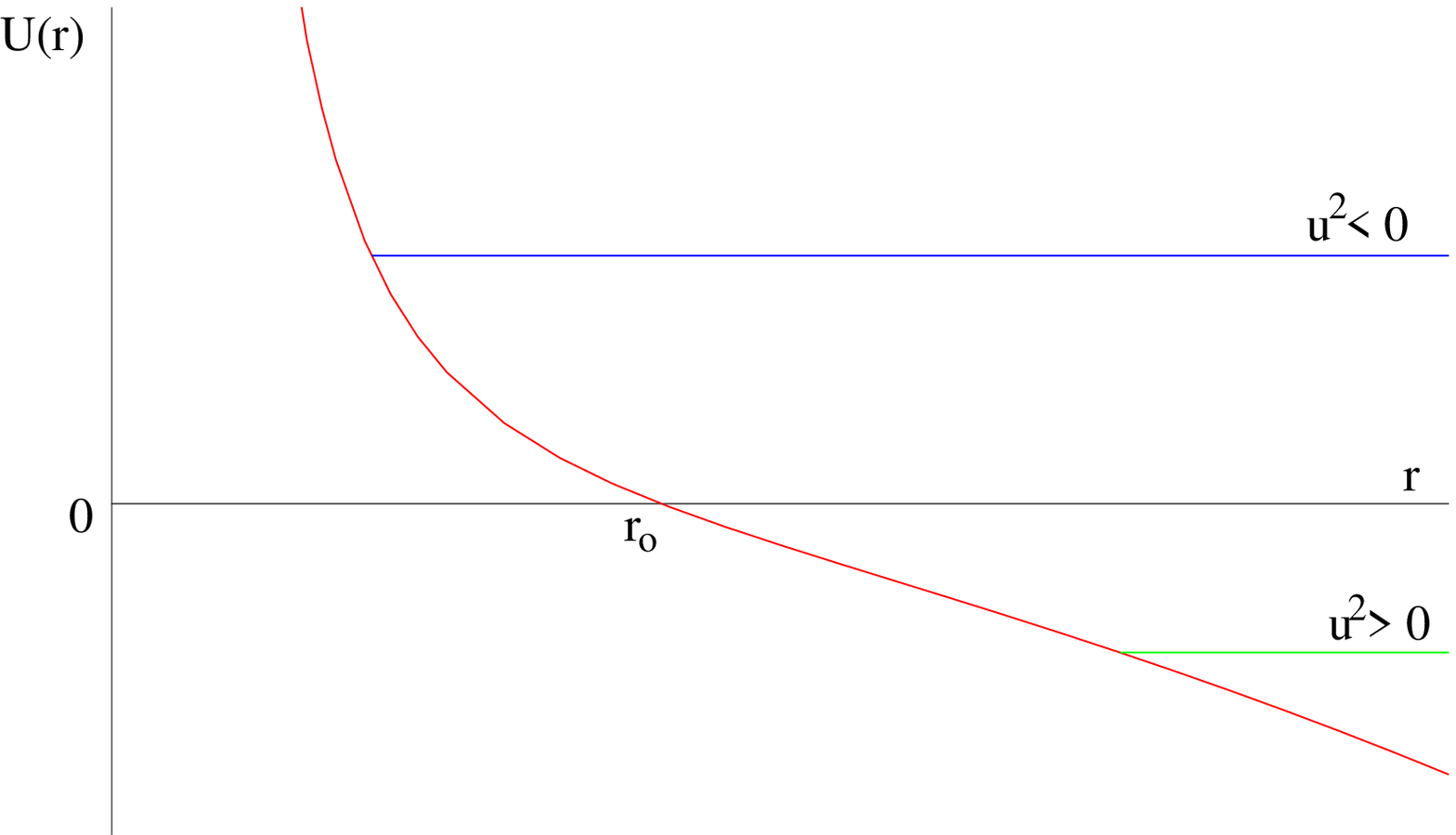}}

With the branch cuts precisely specified,  $r_c (u)$ can now be
uniquely determined from that for $u>0$. In particular, \branchi\
implies that the analytic continuation should be done through the
region around $u=0$. To to be definite, let us concentrate on real
$u^2$. In this case, a convenient way to visualize how the turning
 point changes with $u$ is to treat equation \geodE\ as the motion of a
one-dimensional particle of energy $E^2 = -u^2$, moving in a
potential\foot{Note the potential $V$ is inverted since we work in
the classically forbidden region of the Schrodinger problem
\TeomD.}
 \eqn\effpo{
 U= -V = -f  \ ,
 }
as in \Feffpot. For real $u$, $r_c > r_0$, i.e.
 the turning point lies outside the horizon, while for
 $u$ pure imaginary, $r_c < r_0$ and the the turning point lies inside the horizon.
One can also solve \turnPw\ explicitly and $r_c (u)$ is given by
the positive branch of the equation
 \eqn\Fhsp{
r_c^2 = \sqrt{\mu + \le({1 - u^2 \ov 2} \ri)^2} - {1 - u^2 \ov 2}
\ .
 }

\ifig\geodesicsE{Radial geodesics in the Euclidean section of the
spacetime which corresponds to real values of $u$. The Euclidean
section of the $r-t$ plane is a disk with $i t$ as the angular
coordinate and the origin of the disk at $r=r_0$.  The solid
circle is the boundary. Geodesics with $u>0$ (i and ii) and $u <0$
(iii) are schematically plotted. Geodesic ii correspond to the
large $u$ limit, in which case the tuning point is close to the
boundary and the end points of the geodesic are nearly
coincidental. } {\epsfxsize=3.8cm \epsfbox{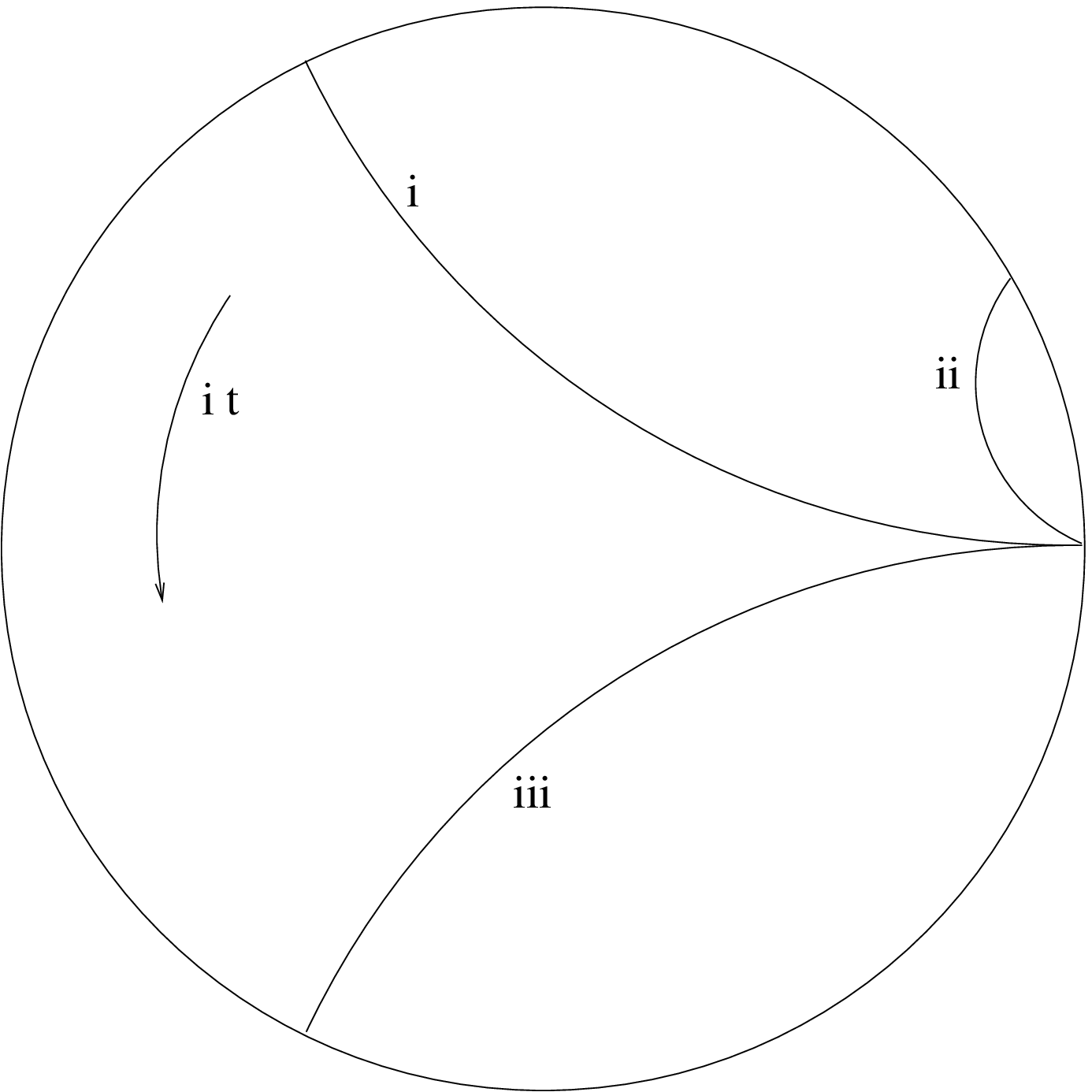}}

\ifig\geodesics{Radial geodesics in the Lorentzian section of the
spacetime corresponding to pure imaginary values of $u = i E$ are
schematically plotted. Geodesics i and ii have $E>0$ while iii has
$E<0$. Geodesic ii correspond to the limit $E \to +\infty$, in
which case the tuning point is close to the singularity and the
geodesic becomes nearly null.} {\epsfxsize=3.8cm
\epsfbox{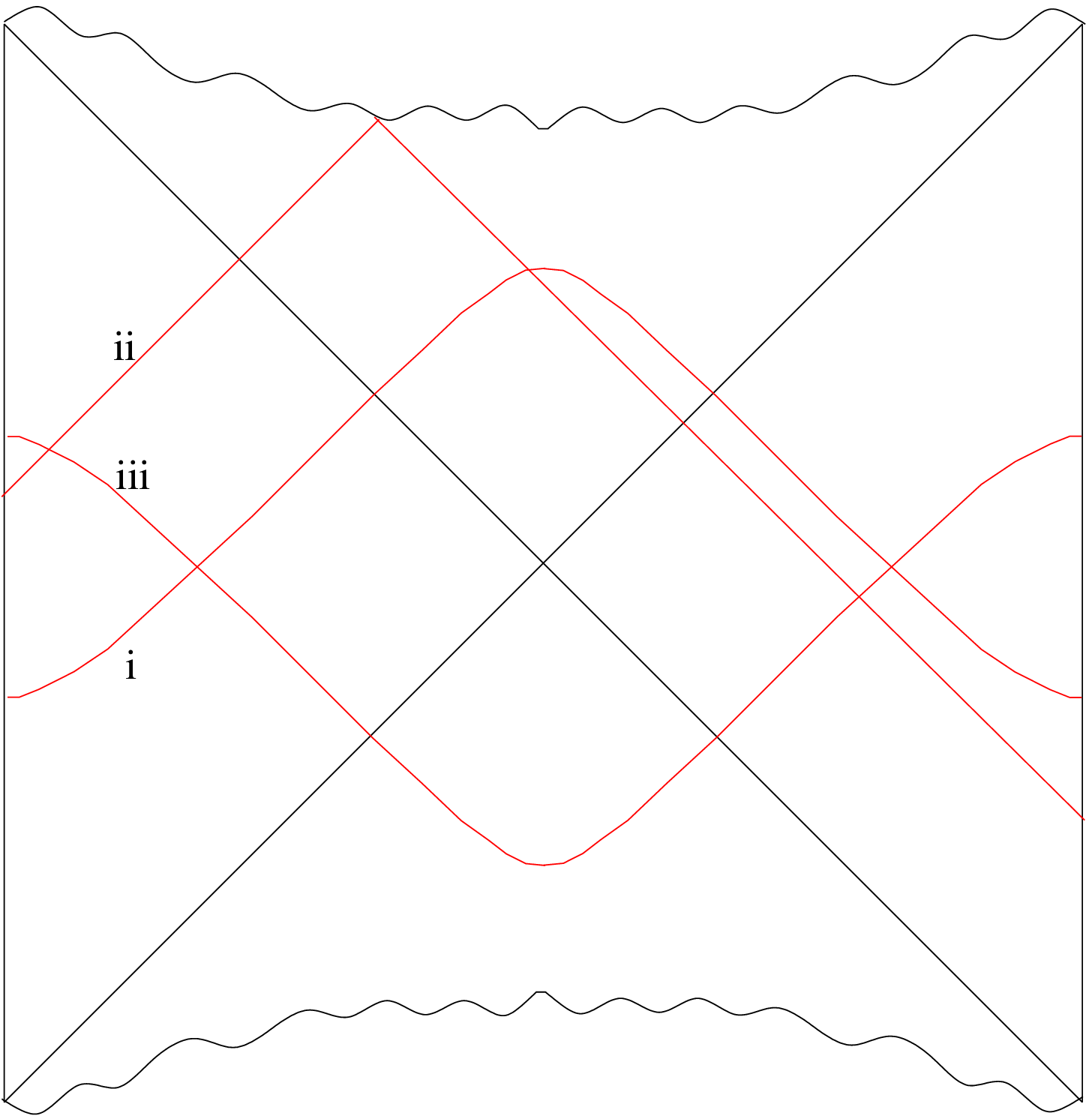}}

With the turning point specified, one can now find the bulk
geodesics corresponding to various values of $u$ from simple
geometric considerations. For example,  real values of $u$
correspond to radial geodesics in the Euclidean section of the
spacetime (see \geodesicsE), since from \geoRe\ and \velocD\ $t$
is pure imaginary along the geodesics. Pure imaginary values of $u
= i E$ ($E$ real) correspond to  real geodesics in the Lorentzian
section of the spacetime, which connect two asymptotic boundaries
(see \geodesics). The turning point of such a geodesic lies in
region II for $E>0$ and in region IV for $E<0$ (see also
\penrose).

The dependence of $r_c$ on $u$ illustrates some interesting
features in the relation between bulk and boundary scales. For
real $u \to \pm \infty$, the turning point is given by
 \eqn\dhsp{
 r_c\approx |u| \to + \infty
  }
i.e. the turning point approaches the boundary. In this limit, the
end points of the geodesic becomes nearly coincidental. When $u$
decreases, $r_c$ also decreases. The turning point $r_c$ reaches
the horizon for $u=0$. This behavior reflects a familiar feature
of the AdS/CFT correspondence, called IR/UV
connection~\refs{\MaldacenaRE,\sussWi}, which relates long
distances in the AdS spacetime to high energies in the boundary
theory. The turning point $r_c$ moves inside the horizon when $u$
moves along the imaginary axis from the origin. Let $ u = i E$.
Then as $|E|$ increases, $r_c$ decreases (see equation \Fhsp). For
$|E| \to + \infty$, we find that
 \eqn\behar{
 r_c \approx {\sqrt{\mu} \ov |E|}  \to 0 \
  }
i.e. the turning point approaches the singularity. Thus when
dealing with physics inside the horizon, there appears to be a new
feature. To probe deeper inside the horizon requires larger $E$.
Since the singularity may heuristically be considered as the UV of
the bulk, we find a UV/UV connection. It is important to keep in
mind that inside the horizon, $r$ plays the role of the time
coordinate. This indicates that the ``time'' inside the horizon is
indeed holographically generated from the boundary Yang-Mills
theory.

The above discussions can be easily generalized to all complex
values of $u$ using equation \turnPw. In particular, from \turnPw,
$|r_c| \to 0$ requires that $|u| \to \infty$, due to the fact that
$f$ blows up at the singularity (large curvature effect).
Conversely, $|u| \to \infty$ implies either $|r_c| \to 0$ or
$|r_c| \to + \infty$. Thus along different directions to infinity
in the complex $u$-plane, the turning point either approaches the
boundary or the singularity. The branch cuts in \branchi\ divide
the complex infinity of the $u$-plane into various asymptotic
regions. The regions which correspond to the singularity or the
boundary are indicated in \branchi.  Near the real $u$ axis, the
turning point approaches the boundary as $|u|
 \to + \infty$.  As $|u| \to + \infty$ near the imaginary $u$
 axis, the turning point approaches the singularity.

\ifig\contour{The integration contours in the complex $r$-plane
for (a): $u>0$, (b): $u<0$, (c): $u = iE, E > 0$, (d): $u=iE,
E<0$. The solid dot indicates the pole $r=r_0$ of the integrand
(horizon). } {\epsfxsize=8cm \epsfbox{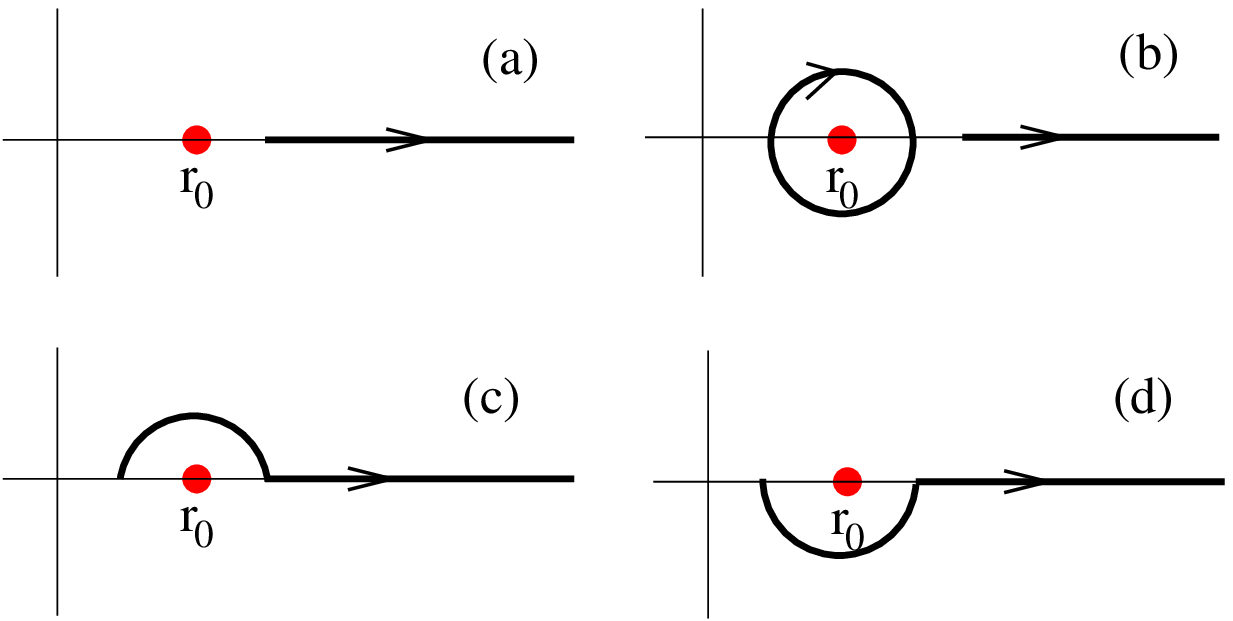}}

We now look at the second aspect of the analytic continuation of
$Z(u)$. This involves the deformation of the integration contour
in \ZdefB\ (or \varSInE) as $r_c (u)$ moves in the complex
$r$-plane. As the contour is deformed, one needs to be careful
about the contribution of the pole at $r=r_0$ of the $1/f$ factor
in the integrand. To find the contribution of the pole, it is
enough to consider when the turning point is close to the
pole\foot{$f$ also has other poles at $-r_0$ and $\pm i r_1$. The
analytic continuation procedure is such that the turning point
never coincides with these poles.}. This happens when $|u|$ is
small, in which case a prescription for the contour deformation
can be obtained from the requirement that $Z(u)$ be analytic at
$u=0$. The contours for other values of $u$ can then be obtained
by continuous deformation without further subtlety. The resulting
contours for real and pure imaginary $u$'s are plotted in
\contour, from which the contribution of the pole can be readily
obtained.

For real $u < 0$, the contribution from the pole to $t(u)$ of
\varSInE\ is $- i \beta$. This is precisely the period of the
complex time and matches with the geometric picture of
\geodesicsE. When $u = iE$ for real $E$, the imaginary part of
$t(u)$ solely comes from the pole contribution and we find
 \eqn\imgTv{
 {\rm Im} \, t (iE) =  - {i \beta \ov 2}, \qquad E \;\; {\rm real}
 \ .
 }
This shows that the end point of the corresponding geodesic lies
in the other asymptotic boundary, i.e. region III of \penrose,
again consistent with \geodesics.

To summarize, through the function $r_c (u)$, we establish a
correspondence between the complex $u$- and $r$-planes. $G_+ (\nu
u)$ evaluated at $u$ in the large $\nu$ limit probes the black
hole geometry near $r_c (u)$. The boundary correlation functions
encode not only the bulk geometry outside the horizon, but also
regions beyond the horizon and near the singularity.

We emphasize that the proper identification of the branch cuts for
$Z(u)$ and $r_c (u)$ (which in turn depends on the knowledge of
poles for $G_+ (\om)$) is crucial for our conclusion above.
Different choices of the branch cuts could lead to completely
different physical pictures. For example, a different analytic
continuation procedure may lead to a $r_c (iE)$ that for real $E$
is not given by the positive branch of \Fhsp, but some other root
of the turning point equation \turnPw. In that case one cannot
associate the geodesics in \geodesics\ with $G_+ (\om)$ and it is
not clear one could probe the physics beyond the horizon.

Given $Z(u)$ from the large $\nu$ limit of the boundary $G_+ (\nu
u)$, the problem of finding the bulk metric is essentially a
classical inverse scattering problem. Due to the large number of
isometries of the background, the problem is effectively
one-dimensional, being that of a particle moving in \effpo. The
equivalent one-dimensional problem can be phrased as follows.
Consider sending a particle toward the potential from $r = \infty$
and waiting for it to come back. $L (E)$ is then the (regularized)
time interval for this scattering process. With the knowledge of
$L (E)$ for all values of $E$, one can in principle reconstruct
the potential \effpo. At a given $E$, $L (E)$ probes the behavior
of the potential \effpo\ near the turning point $r_c (u)$.

\subsec{Manifestations of singularities in boundary theories}

We have found that the geometry around the black hole singularity
is encoded in the behavior of $G_+ (\om)$ near the imaginary
infinity. We now examine the manifestations of the singularity
explicitly.

The integrals in \varSInE\ can be evaluated explicitly and one
finds\foot{The expressions below were obtained before
in~\refs{\shenker} in a somewhat different form.}
 \eqn\defZz{\eqalign{
 L(u) & = - \ha \log (A_+ \tilde A_+ A_- \tilde A_-)  + 2 \log {|B| \ov 2 \pi} \cr
 t(u) & =  {\beta \ov 4 \pi} \log \le({A_+ \tilde A_- \ov A_- \tilde A_+} \ri)
  - i{\tilde \beta \ov 4 \pi} \log \le({A_+ \tilde A_+ \ov A_- \tilde A_-}
  \ri) - {i \beta \ov 2} \cr
 }}
and
 \eqn\adeff{
A_\pm = \ha \pm \rw {\cB \ov 4 \pi}, \qquad \tilde A_\pm = \ha \pm
\rw {\bar \cB \ov 4 \pi}
 }
In \defZz\ the branch cuts of the logarithms are chosen to be
straight lines extending radially from $\pm {2 \pi \ov \cB}$ and
$\pm {2 \pi \ov \bar \cB} $ to $\infty$, as follows from the
discussion in previous subsection (see \branchi).

 Expanding \defZz\ for large $|u|$ we
find that $L(u)$ and $t(u)$ of \varSInE\ can be written as
 \eqn\gsjrB{\eqalign{
 L(u) & = - 2 \log \le({[u] \ov 2}\ri) + \sum_{n=1}^\infty {a_{2n} \ov 2n} {1 \ov u^{2n}}   \cr
 t(u) & = t_0  - i{2 \ov  u} - i \sum_{n=1}^\infty {a_{2n} \ov 2n+1}
 {1 \ov u^{2n+1}}\cr
  }}
which lead to the expansion for $Z(u)$
 \eqn\larEb{\eqalign{
  Z (u)& = i u t_0  +  2 \log  {[u] \ov 2}  + 2 - \sum_{n=1}^\infty {a_{2n} \ov 2n (2n+1)}
  {1 \ov u^{2n}}  \ .  \cr
 }}

In \gsjrB\ and \larEb,
$$
a_n = \le({2 \pi \ov \cB}\ri)^n + \le({2 \pi \ov \bar \cB}\ri)^n \
$$
and $t_0$ and $[u]$ are given by
 \eqn\thave{
t_0 = \cases{ 0   & $u \in  \; B_+$ \cr\cr
              - i \beta  & $u \in  \; B_-$ \cr\cr
              {\bar \cB \ov 2} & $ u \in  \; S_+$ \cr\cr
             - { \cB \ov 2} &  $ u \in  \;  S_-$
             }  , \qquad
 [u] = \cases{ u   & $u \in B_+ $ \cr\cr
              -  u   & $u \in B_- $ \cr\cr
              - i u & $ u \in S_+$ \cr\cr
             i u  &  $ u \in S_-$
                  }\ ,
             }
where $B_\pm$ and $S_\pm$ denote asymptotic regions in \branchi\
whose corresponding turning point approaches the boundary and the
singularity respectively. The values of $t_0$ for various limits
can be easily understood from the geometric pictures of the
geodesics in \geodesicsE\ and \geodesics. When $u \to \pm \infty$,
the end points of the Euclidean geodesics become nearly
coincidental. The value $- i \beta$ for $u \to -\infty$ is
precisely the full period of the Euclidean circle. When $u \to \pm
i \infty$, the bouncing geodesics in \geodesics\ become nearly
null and the corresponding values of $t_0$ in \thave\ are twice of
those in \geTi.

 Equations \larEb\ implies that as $\om = \nu u \to \pm i \infty$, the boundary
correlation function behaves as
 \eqn\alongL{\eqalign{
 G_+ (\om, l=0)
  &  \approx {1 \ov \pi (\Ga (\nu))^2} \le(\mp i {\om  \ov 2} \ri)^{2 \nu}  \,
   e^{  i \om  \le( \pm {\tilde \beta \ov 2}
 - {i \beta \ov 2} \ri) } \le(1 + O\le({1 \ov \om^2}\ri)\ri)  \cr
 }}
where the upper (lower) sign corresponds to $\om \to + i \infty$
($-i \infty$). Note that the correlation function decays
exponentially along these directions. For $\om \to \pm \infty$
near the real axis, we find
 \eqn\norcon{\eqalign{
 G_+ (\om, l=0) & \approx  \cases{
 {1 \ov \pi (\Ga (\nu))^2} \,
 \le({\om  \ov 2} \ri)^{2 \nu}  \le(1 + O\le({1 \ov \om^2}\ri)\ri)
 & $  \om \to + \infty$ \cr\cr
   {1 \ov \pi (\Ga (\nu))^2} \,
 \le(-{\om  \ov 2} \ri)^{2 \nu} \,
 e^{\beta \om }  \le(1 + O\le({1 \ov \om^2}\ri)\ri)
 & $  \om \to - \infty$ \cr
 } \ .
 }}
While equations \alongL\ and \norcon\ were derived in the large
$\nu$ limit, they should hold for finite $\nu$, since the $|u| \to
\infty$ limit should coincide with the limit $|\om| = \nu |u| \to
\infty$ regardless of the value of $\nu$. Note that \norcon\ is
precisely what one would expect of the large frequency behavior of
a conformal field theory at finite temperature\foot{When real $\om
\to + \infty$, one expects the correlation function to recover the
zero temperature result. The second line of equation \norcon\
follows from the general properties of the Wightman function at
finite temperature.}. The exponential falloff in \alongL\ reflects
the presence of a curvature singularity in the bulk. The falloff
is controlled by the complex parameter $\cB$ (introduced in
\defBB) which characterizes the black hole geometry.

\subsec{Generalizations to nonzero angular momentum}

The above discussions can be extended to Wightman functions of
nonzero angular momenta. We summarize some main results here,
leaving detailed discussions to~\refs{\FLlong}:

\item{1.} For boundary angular velocity $k$ real, which
corresponds to geodesics of pure imaginary angular momentum, the
structure of the branch cuts for $Z(u,k)$ is similar to \branchi.
The locations of the branch points and the directions of the
branch cuts depend nontrivially on $k$. At finite $\nu$, the
branch cuts become lines of poles of $G_+ (\om, l)$.

\item{2.}For bulk geodesics with real angular momentum $q$,
  an important new feature appears: there exist geodesic orbits
with constant real $r$. The existence of such orbits leads to the
appearance of virtual states (if there is an orbit lying inside
the horizon) or bound states (if there is an orbit lying outside
the horizon) in the Schrodinger problem \TeomD. These virtual
states or bound states lead to two new lines of poles of $G_+
(\om,l)$ along the imaginary $\om$-axis for pure imaginary $l$.
Thus $Z(u,k)$ has two new branch cuts along the imaginary $u$-axis
for $k = i q$ pure imaginary.

\item{3.} It remains true that the turning point approaches the
boundary (singularity) for $|u| \to \infty$ along the real
(imaginary) axis. Furthermore, for any fixed $l$, in the large
$\om$ limit, equations \alongL--\norcon\ remain valid.

\item{4.} In the limit that $q$ goes to zero, the branch cuts for
$Z(u,iq)$ along the imaginary axis move to infinity and \branchi\
is recovered. The fact that there are branch points at $u = \pm i
\infty$ at $q=0$ leads to interesting behaviors in the expansion
of $Z(u,k)$ around $k=0$. For example, let $u = i E$ with real
$E$, then one finds that to leading order in the limit $E \to +
\infty$, $Z(u,k)$ has the following small $k$ expansion\foot{Note
that the equation below is obtained by first doing a small $k$
expansion and then taking leading order terms in $1/E$.}
 \eqn\adosn{
 Z (iE, k) \approx - E {\bar \cB \ov 2} + 2 \log  {E \ov 2} +2 + {1 \ov E^2}
 \sum_{l=1}^\infty a_l \le( k^2 E^2
\ri)^l + \cdots
 }
 with
 \eqn\parMe{
 a_1 = - {\pi \ov 2 \mu^\ha }, \qquad a_2 = {3 \pi \ov 16 \mu^{3 \ov
 2}}, \qquad a_l \sim {1 \ov \mu^{2l-1 \ov 2}} \
 }
where $\mu$ dependence can be deduced based on dimensional
analysis. Note that the expansion parameter for small $k$ is given
by $k^2 E^2$ and the derivatives over $k$ at $k=0$ become
divergent in the large $E$ limit. We will see in the next section
that \adosn\ leads to divergences in certain gauge invariant
observables in the boundary theory.

\ndt We also mention by passing a few other generalizations:

\item{5.} The discussions of this section can also be generalized
to an AdS$_{d+1}$ black hole of dimension $d \geq
2$~\refs{\FLlong}. All the essential features for AdS$_5$ black
holes carry over to other dimension $d \geq 3$. A special case is
a BTZ black hole in AdS$_3$, which has an orbifold singularity and
the story is somewhat different. In the BTZ case, the
corresponding Schrodinger equation \TeomD\ can be exactly solved
and the relation between the large $\nu$ limit of Wightman
functions and bulk geodesics can be explicitly verified.

\item{6.} The subdominant contributions in \bdgpE\ can also be
worked out using a more sophisticated WKB method involving more
than one turning point. One can show that an infinite number of
subdominant contributions become important at the branch
cuts\foot{In other words, the branch cuts are anti-Stokes lines
for an infinite number of subdominant exponentials. The Stokes
lines can be obtained from the branch cuts by rotating $\pm {\pi
\ov 2}$. A simple example which also exhibits this phenomenon is
the Gamma-function (see e.g.~\refs{\berryG}).} where they add up
to produce the poles of $G_+ (\om,l)$. In~\refs{\FLii} we use this
property to derive the positions of poles of $G_+$ for general $l$
and dimension in the large $\nu$ limit.

\item{7.} While we have not examined it in detail, it is
interesting to compute the higher order $1/\nu$ corrections in
\bdgpE. In particular, the function $Q(z)$ \liuoG\ will start
contributing at the order $O(1/\nu)$. Since $Q(z)$ becomes
singular at $r \to 0$, it would be interesting to see whether it
yields new manifestations of the singularity in the boundary
theory correlation functions.

\subsec{Coordinate space correlators and alternative indications
of curvature singularities}

In this subsection we consider the Fourier transform of $G_+ (\om,
l)$ to coordinate space. To make connection to the result
of~\refs{\shenker}, we consider
 \eqn\realT{\eqalign{
 G_{12} (t, \th=0) & = G_+ \le(t - {i \beta \ov 2}, \th=0 \ri) \cr
  }}
which can be obtained from \GfouB\ by taking $t \to t - {i \beta
\ov 2}$ and corresponds to inserting operators on two different
boundaries. We restrict to $\th =0$ for simplicity. In the large
$\nu$ limit, $G_{12}$ can be evaluated in exact parallel of the
discussion of section 3.3. One can approximate the sum over $l$ by
an integral
 \eqn\sdaInt{\eqalign{
 G_{12} (t) & \approx   {\nu^4 \ov 8 \pi^3}
 \int_{-\infty}^\infty  d u  d  k \, k^2\,  e^{-i \nu u t - \ha \nu u \beta} \,
 2 \nu e^{\nu Z (u, k)}  \cr
 }}
and perform the integrals using the saddle point method.

\ifig\contour{The contour plot for the real part of $Z(u,k=0)-i u
(t- i {\beta \ov 2})$ for $t < {\tilde \beta \ov 2}$ in the
complex $u$-plane. There is a saddle on the imaginary axis and two
complex ones. The steepest descent contour is also shown in
figure. The contour does not pass through the saddle on the
imaginary axis even if it dominates.} {\epsfxsize=6cm
\epsfbox{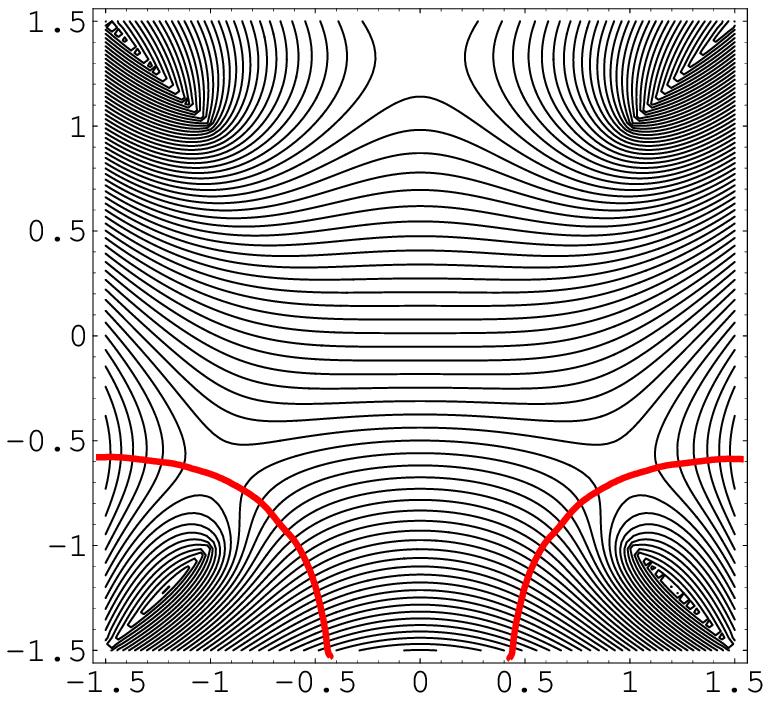}}

From the discussion of section 3.3, the saddles of \sdaInt\
correspond to geodesics whose end points lie on different
boundaries and are separated in time by an amount $t$ and no
separation on $S^3$. The saddle for the $k$-integral is simply
$k=0$. The saddles for the $u$-integral are given by
 \eqn\sdal{
{\p Z (u,k=0) \ov \p u} = i t + {\beta \ov 2} \ .
 }
Note that $Z(u,k=0)$ can be obtained from \defZz. The solutions to
\sdal\ can be visualized conveniently on the contour plot of the
real part of $Z(u,k=0)-i u (t- i {\beta \ov 2})$ in the complex
$u$-plane. See \contour. In the figure we also indicate the
steepest descent paths to which the integration contour of
\sdaInt\ can be deformed.

The dependence of the saddle point structure on $t$ can be
summarized as follows. For $t< t_c ={\tilde \beta \ov 2}$, there
are three saddles, as indicated in \contour. The one on the
imaginary axis corresponds to a real geodesic in Lorentzian black
hole spacetime with a turning point inside the horizon (see
\geodesics). This is the bouncing geodesic discussed
by~\refs{\shenker}. We will refer to this saddle as the bouncing
saddle below. The other two saddles describe complex geodesics,
which do not seem to probe the physics beyond the horizon. As $t$
approaches $t_c$, the bouncing saddle moves to infinity along the
imaginary axis and the turning point of the geodesic approaches
the singularity.  For $t
> t_c$, the bouncing saddle disappears\foot{For $t > t_c$, the two complex saddles
wind around the branch points at ${2 \pi \ov \cB}$ and $-{2 \pi
\ov \bar \cB}$ respectively and approach them as $t \to \infty$.}.

 From the
steepest descent contour, we conclude that the bouncing geodesic
does not contribute to coordinate space correlation functions.
This result was obtained in~\refs{\shenker} by analytical
continuation from Euclidean signature\foot{In~\refs{\shenker} it
was argued that the information regarding the bouncing geodesics
and thus the singularity can nevertheless be obtained by analytic
continuation in the large $\nu$ limit, since there exist certain
values of $t$ for which the bouncing saddle merges with other
saddles.}. Here we confirm their result.

\subsec{New observables of the boundary theory and manifestations
of the curvature singularity}

While the bouncing geodesic does not contribute to \realT\
directly, from momentum space correlation functions, we can easily
construct observables in the boundary theory which are
approximated by the bouncing geodesics in the large $\nu$ limit.

\ifig\Ucontour{The integration contours for $G_{12} (t)$ ($\CC_1$)
and $H_{12} (\tau)$ ($\CC_2$).} {\epsfxsize=6cm
\epsfbox{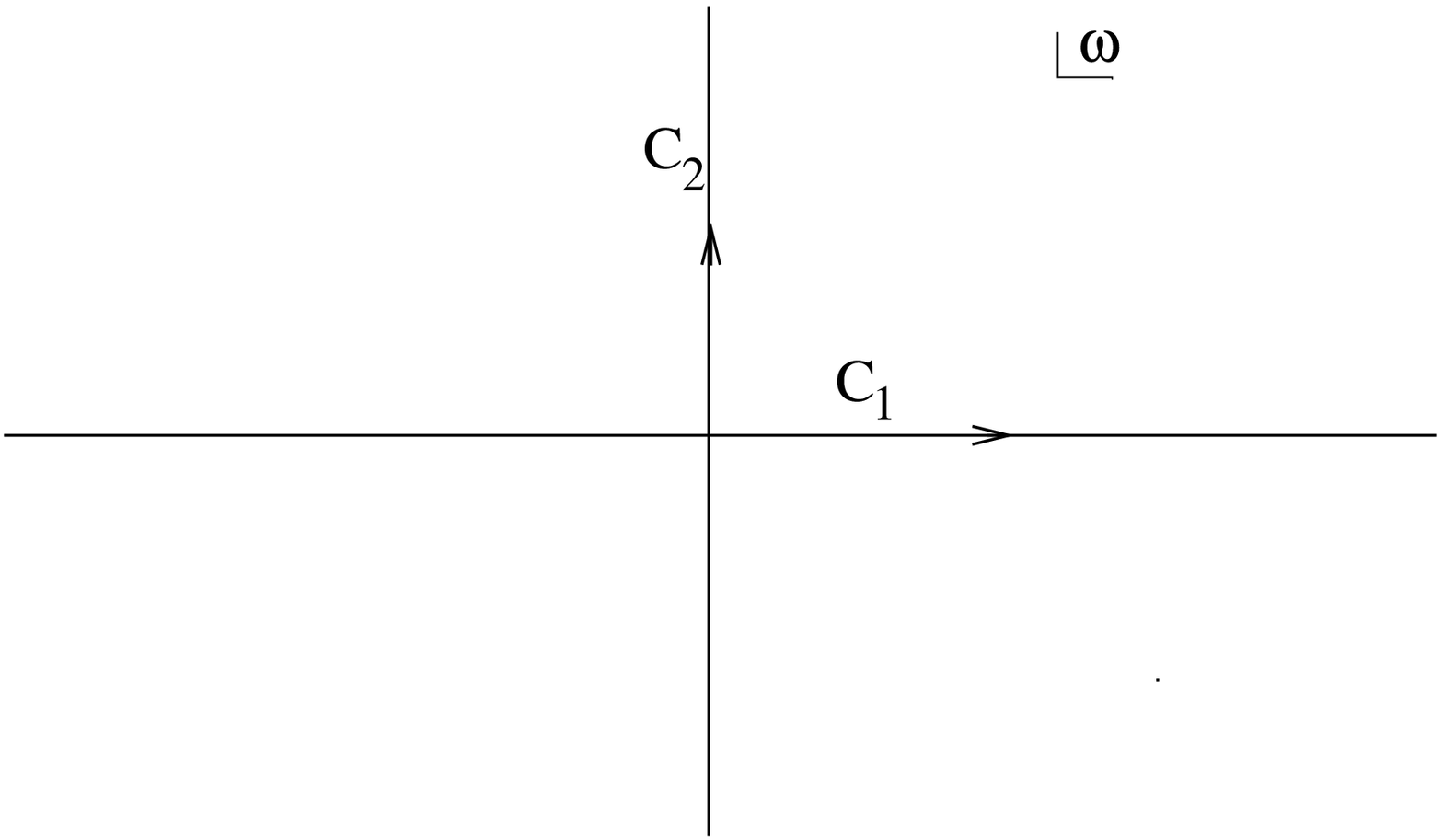}}

Let us start with two-sided correlator \realT\ with coincidental
spatial coordinates
 \eqn\Gont{\eqalign{
 G_{12} (t) & = \Tr \le[e^{-\beta H} \OO (t-i\beta/2,e) \OO (0,e)
 \ri] \cr
 & = {1 \ov 4 \pi^2} \sum_{l=0}^\infty 2 (l+1)^2  \int_{\CC_1}  {d\om
 \ov 2 \pi} \, e^{-i \om t} G_{12} (\om,l)
 }}
where the contour $\CC_1$ is along the real $\om$-axis and
 \eqn\cealRm{\eqalign{
 G_{12} (\om, l)
  = e^{-{\om \beta \ov 2}} G_+ (\om,l) \ .
 }}
From \alongL-\norcon\ and the comments in sec. 4.3 regarding their
generalizations to nonzero $l$, we find that the large $\om$
behaviors of $G_{12} (\om,l)$ are given by
 \eqn\gianii{\eqalign{
 G_{12} (\om, l)
 \approx  \cases{
    {1 \ov \pi (\Ga (\nu))^2} \le(\mp i {\om  \ov 2} \ri)^{2 \nu}  \,
   e^{ \pm  {i \om   \tilde \beta \ov 2}}
  & $\om \to \pm i \infty$
 \cr\cr
 {1 \ov \pi (\Ga (\nu))^2} \,
 \le(\pm {\om  \ov 2} \ri)^{2 \nu} e^{\mp {\beta \om \ov 2}}
 & $  \om \to \pm \infty$ \cr\cr
}
 }}
We now construct new observables in the boundary theory
 \eqn\newOn{
 H_{12} (\tau) = {1 \ov 4 \pi^2} \sum_{l=0}^\infty 2 (l+1)^2  \int_{\CC_2}  {d\om
 \ov 2 \pi} \, e^{-i \om \tau} G_{12} (\om,l)
 }
where the contour $\CC_2$ are along the imaginary axis. $H_{12}
(\tau)$ can be defined for $-{\tilde \beta \ov 2} < \tau < {\tilde
\beta \ov 2}$ due to the exponential falloff \gianii\ of $G_{12}
(\om)$ along the imaginary axis. Note that $H_{12} (\tau)$ is
gauge invariant by definition. While its existence depends on the
asymptotic behavior of $G_{12} (\om,l)$ along the imaginary
$\om$-axis, it is an object which can in principle be
intrinsically defined in the boundary theory\foot{One can of
course define $H_{12} (\tau;e,e')$ for non-coincidental points on
$S^3$. We do not expect divergences for them. They will be
discussed in~\refs{\FLlong}.}.

In the large $\nu$ limit \newOn\ can be evaluated in exact
parallel as \realT--\sdaInt\ by approximating the sum over $l$ by
an integral and performing the saddle point approximation in the
resulting integrals. The only difference is that the integration
contour for $\om$ is now along the imaginary axis rather than the
real axis. Thus instead of picking up the two complex saddles in
\contour,
\newOn\ is given by  expanding around the bouncing
saddle on the imaginary axis.

As $\delta t = t_c - \tau  \to 0$, the bouncing saddle moves to
the imaginary infinity and the turning point $r_c$ of the
corresponding geodesic approaches the singularity. In this limit,
we can use \adosn\ to approximate $Z(u,k)$. Then the integrals for
$H_{12}$ can be written as ($u= iE$)
  \eqn\divgP{\eqalign{
 H_{12} (\tau) & \sim
 \int dE \int d k k^2 \; \exp \le[- \nu E \delta t + 2 \nu \log
{E \ov 2} + {\nu \ov E^2} \sum_{l=1}^\infty a_l \le( k^2 E^2
\ri)^l \ri] \cr
 & \sim (E_c)^{2 \nu +1} \le(1 + \sum_{n=1}^\infty c_n \le({E_c^2
\ov \nu \sqrt{\mu}}\ri)^n \ri) \cr
 & \sim  {1 \ov
 (\delta t)^{2 \nu + 1}}  \le(
   1 + \sum_{n=1}^\infty {b_n \ov
 (\nu \sqrt{\mu} \delta t^2)^{n}} \ri) \cr
 }}
where  the saddle for the $k$-integral is $k=0$ and for the $E$
integral is
 \eqn\Esdal{
E_c \approx {2 \ov \delta t} \to \infty \ .
 }
Note that in \divgP\ we have only worked out the qualitative
$\delta t$ dependence in the small $\delta t$ limit rather than
attempting a precise evaluation. The order of limits should be
first $\nu \to \infty$ and then $\delta t \to 0$.  In evaluating
the high order $1/\nu$ terms in \divgP, there are two other
sources of $1/\nu$ corrections that we did not take into account:
$1/\nu$ corrections in \bdgpE\ and those in turning the sum over
$l$ in
\newOn\ to an integral.  Since we are only interested in power
counting, apart from magic cancellations, this will not affect the
qualitative behaviors in~\divgP. The divergences in $1/\nu^n$
terms  arise from the $k$-integral and are due to the structure of
the small $k$ expansion in \adosn.

Note that \divgP\ is precisely what one expects of a bouncing
geodesic as its turning point approaches the singularity. From the
bulk point of view, one expects the contribution from the geodesic
 should take the form\foot{Similar argument was also used in~\refs{\shenker}.}
  \eqn\BDCo{\eqalign{
    J^\ha e^{-\nu L} \le(1 + \sum_{n=1}^\infty
 {R_n \ov \nu^n} \ri)
 }}
where $L$ is the regularized the geodesic distance,  $J$ is the
Van Vleck-Morette determinant. Higher order terms in \BDCo\ arise
from cubic and higher order terms in the sum over paths around the
geodesic and $R_n$ can be expressed in terms of bulk geometric
quantities in the form of components of curvature tensors (and
their derivatives) integrated along the geodesic. In general the
explicit expressions for $R_n$ are very complicated~(see
e.g.~\refs{\bekenstein}). On dimensional ground, one expects that
in the limit that the turning point approaches the singularity
$R_n \sim {1 \ov \ep^n}$, with $\ep$ the proper time between the
turning point and the singularity. This  leads to
 \eqn\BDCom{\eqalign{
   H_{12} \sim  J^\ha e^{-\nu L} \le(1 + \sum_{n=1}^\infty
 {b_n \ov (\nu \ep)^n} \ri) \ .
 }}
Equation \BDCom\ precisely agrees with \divgP\ since from the
metric \defF\
$$
\ep \sim  \int_0^{r_c} {dr \ov \sqrt{f}} \sim {r_c^2 \ov
\sqrt{\mu} } \sim {\sqrt{\mu} \ov E_c^2} \sim \sqrt{\mu} \delta
t^2
$$
where we have used \behar\ and \Esdal.

To summarize we have constructed  new gauge invariant observables
\newOn\ in the boundary theory that are sensitive to the physics
beyond the horizon and in particular their behaviors near $\tau
\to {\tilde \beta \ov 2}$ precisely reflect the curvature
divergence of the singularity. These observables are related
nonlocally in time with \realT. It would be interesting to better
understand their meaning in Yang-Mills theories\foot{It seems
possible to define a new theory whose momentum space correlation
functions evaluated at $\om$ are given by $G_{+} (i \om)$. $H_{12}
(\tau)$ would correspond to ``Euclidean'' Green functions in this
new theory.}.

\newsec{Discussions: Resolution of black hole singularities at finite $N$ ?}

In this paper we established a direct relation between space-like
geodesics in the bulk and the large operator dimension limit of
the boundary Wightman functions $G_+ (\om, l)$  in momentum space.
The results present an intriguing picture on how physics beyond
the horizon is encoded in the boundary theory. In particular, it
gives a clear indication that the ``time'' inside the horizon is
holographically generated from the thermal Yang-Mills theory.

The poles of  $G_+ (\om, l)$
 separate the asymptotic region of the complex $\om$-plane into several
 sectors (see \poles\ and \branchi).
 The sectors near the real axis describe the
 physics near the boundary while the sectors near the imaginary
 axis describe the physics near the singularity\foot{
 We note that this conclusion does not depend sensitively on whether
  the curvature singularity lies in the Lorentzian section of spacetime.
 For example, if one regularizes $f$ in \defF\ by $f = r^2 +1 - {\mu \ov r^2 +
 \ep^2}$ with a small $\ep$ to move to the singularity to the complex
 $r$-plane~\refs{\KaplanQE}, the
 picture would remain exactly the same, provided this does not change the pattern of
 quasinormal poles significantly, which should be the case for $\ep$ small.}.
We found the following signals of the singularity in the boundary
theory:

\item{1.} $G_+ (\om,l)$  falls off  exponentially for $\om \to \pm
i \infty$ (see equation~\alongL\ or \gianii). The falloff is
controlled by the complex parameter $\cB$ \defBB, which
characterizes the black hole geometry.

\item{2.} We constructed new observables $H_{12} (\tau)$ (equation
\newOn) in the boundary theory which are related nonlocally in
time with coordinate space Wightman functions. The curvature
divergence of the singularity is reflected in the divergences of
$H_{12} (\tau)$ as $\tau \to \pm { \tilde \beta \ov 2}$. While the
leading order divergence of $H_{12} (\tau)$ (i.e. the prefactor in
\divgP) can be attributed to 1. above, the divergences in higher
order $1/\nu$ terms are due to more delicate behavior of $Z(u,k)$
for small $k$ and $u \to \pm i \infty$.

\ndt While the above results were derived in the large $\nu$
expansion, the essential features should persist for finite $\nu$.
For example, we expect equations \alongL\ and \adosn\ should be
valid for finite $\nu$ as well\foot{This can be checked explicitly
using other approximations.}.

We now comment on the implications of our results in resolving the
black hole singularity.

The rich analytic behavior observed for $G_+$ in the complex
$\om$-plane is tied to the fact that in the large $N$ limit, the
boundary theory has a continuous spectrum, even though it lives on
a compact space. In the bulk, the continuous spectrum arises
because of the presence of the horizon.  In Yang-Mills theory, the
continuous spectrum should be related to the fact that in the high
temperature phase, typical states in the thermal ensemble have
energy of order $N^2$.\foot{These states have degeneracies of
order $e^{c N^2}$ for some constant $c$ in free theory. When
turning on interactions, one expects the degeneracy is lifted and
the energy levels have spacings of order $e^{-cN^2}$. This gives
rise to a continuous spectrum in the infinite $N$ limit. We thank
O.~Aharony, M.~Douglas and S.~Minwalla for a discussion of this
point. Also note that in the low temperature phase, the Yang-Mills
theory has a discrete spectrum even at $N = \infty$.}

At finite $N$, no matter how large, the boundary theory on $S^{3}$
has a discrete spectrum. In particular, the finite temperature
Wightman function should have the form
 \eqn\figdn{
 G_+ (\om) = 2 \pi \sum_{m,n} e^{-\beta E_m} \rho_{mn} \delta
 (\om- E_n + E_m)
 }
which is a sum of delta functions along the real $\om$-axis, where
$m,n$ sum over the physical states of the theory. $G_+ (\om)$ in
equation \figdn\ does not have an unambiguous continuation off the
real axis.  In particular, the procedures of analytically
continuing $G_+$ to complex $\om$ and taking the large $N$ limit
do not commute. Equation \alongL\ arises by taking the large $N$
limit first and then doing the analytic continuation. This appears
to imply that at finite $N$, geometric notions associated with a
black hole, such as the event horizon and the singularity, no
longer exist. This is not surprising since the black hole geometry
arises as a saddle point in the path integral of the boundary
theory in a $1/N$ expansion. If one does not use such an
expansion, the geometric notions lose their meaning. Thus the
singularity appears to be resolved at finite $N$.

The above arguments, however, do not tell us how the singularity
is resolved. There are several possibilities according to which
the singularity can be resolved in AdS/CFT:

\item{I.}  The singularity is already resolved by $\apr$-effects
in perturbative string theory, i.e. at finite 't Hooft coupling
and infinite $N$.

\item{IIa.} The singularity is resolved {\it only} at finite $N$.
But at infinite $N$ there is a large $N$ phase transition at
certain value of the 't Hooft coupling.

\item{IIb.} The singularity is resolved {\it only} at finite $N$
and there is no large $N$ phase transition for any 't Hooft
coupling.

\ndt To see which possibility is realized, it is important to
investigate whether the signals of the singularity found this
paper and in~\refs{\shenker} persist to weak coupling in the large
$N$ limit. If the answer is yes, it would strongly suggest
possibility IIb above. This would be a very desirable situation
since one would then be able to study the black hole singularity
in string theory using {\it weakly} coupled Yang-Mills theory and
focusing on the large $N$ limit. If the answer is no, then both I
and IIa are possible. In the event that IIa is realized, one
should still be able to detect signals of the singularity at weak
coupling, even though the precise signals may not be directly
obtainable from the results at strong coupling.

In any case, we believe the results in the paper should provide a
valuable guide for understanding the black hole singularity in
AdS/CFT.

Finally, it would be interesting to apply the techniques we
developed in this paper to other backgrounds, like charged or
rotating black holes.

\bigskip
\noindent{\bf Acknowledgments}

We would like to thank O.~Aharony, M.~Douglas, Q.~Ejaz,
D.~Freedman, G.~Horowitz, R.~Jaffe, M.~Kruczenski, S.~Mathur,
J.~Maldacena, S.~Minwalla, J.~Negele, K.~Rajagopal, M.~Rozali,
A.~Scardicchio, R.~Schiappa, N.~Seiberg, A.~Sen, S.~Shenker,
D.~Son, L.~Susskind, W.~Taylor, B.~Wecht and B.~Zwiebach for very
useful discussions. We also want to thank Center of Mathematical
Sciences at Zhejiang university for hospitality during part of the
work. This work is supported in part by Alfred~P.~Sloan Foundation
and funds provided by the U.S. Department of Energy (D.O.E) under
cooperative research agreement \#DF-FC02-94ER40818.

\appendix{A}{Fourier transform on $S^3$}

A complete set of scalar harmonics on $S^3$ can be written as $
Y_{lm m'} (e) $ transforming under $(l/2, l/2)$ representations of
$SO(4) = SU(2) \times SU(2)$ with $-l/2 \leq m, m' \leq l/2$. We
use $e$ to denote a point on $S^3$ and $I = (l,m,m')$ to denote
the full set of indices. $Y_I$ is normalized  so that
$$\eqalign{
 & \int_{S^3} Y^I (e) Y^J (e) = \delta_{IJ} \cr
 & \sum_{m,m'} Y_{I} (e_1) Y_{I} (e_2) = {1 \ov 4 \pi^2} { 2 (l+1)}
C_l (e_1 \cdot e_2)
 }$$
with
$$
C_l (\cos \th) = {\sin (l+1) \th \ov \sin \th}
$$
Also note that
$$\eqalign{
  & \nabla^2_{S^3} Y^I  = - l (l+2)  { Y^I}
 }$$
where $\nabla^2_{S^3}$ is the Laplace operator on $S^3$.

Consider a correlation function $G(e,e')$ on $S^3$ which only
depends on the geodesic distance between two points $e$ and $e'$.
It can be expanded as
$$\eqalign{
 G(e,e')  & = {1 \ov 4 \pi^2} \sum_{l=0}^\infty  { 2
(l+1)}  C_l (e \cdot e') G (l)   \cr
 & = \sum_I
\, G (l) \, Y^I (e) Y^I (e')  \ . \cr
 }$$

\listrefs

\end
\end